\documentclass[11pt]{article}

\newcommand{\fft}[2]{{\frac{#1}{#2}}}
\newcommand{\ft}[2]{{\textstyle{\frac{#1}{#2}}}}
\newcommand{\sqr}[2]{{\vcenter{\vbox{\hrule height.#2pt
        \hbox{\vrule width.#2pt height#1pt \kern#1pt
        \vrule width.#2pt}\hrule height.#2pt}}}}

\newcommand{\be}{\begin{equation}}
\newcommand{\ee}{\end{equation}}
\newcommand{\bea}{\begin{eqnarray}}
\newcommand{\eea}{\end{eqnarray}}
\newcommand{\nn}{\nonumber}
\def\1{e^{\frac1{\sqrt6}\phi}}
\def\2{e^{\frac2{\sqrt6}\phi}}
\def\no{e^{-\frac1{\sqrt6}\phi}}
\def\nt{e^{-\frac2{\sqrt6}\phi}}
\def\4{e^{\frac4{\sqrt 6}\phi}}
\def\nf{e^{-\frac4{\sqrt 6}\phi}}

\usepackage{amsfonts}
\usepackage{graphicx}
\usepackage[nosort]{cite}

\def\a{\alpha}

\def\b{\beta}

\def\d{\delta}

\def\ep{\epsilon}

\def\f{\frac}
\def\g{\gamma}
\def\G{\Gamma}

\def\na{\nabla}
\def\nn{\nonumber}

\def\p{\partial}
\def\P{\Phi}

\def\s{\sigma}

\def\w{\wedge}

\def\bep{\bar{\epsilon}}

\makeatletter \@addtoreset{equation}{section} \makeatother

\begin{document}

\begin{titlepage}

\leftline{\tt hep-th/0605268}

\vskip -.8cm

\rightline{\small{\tt MCTP-06-10}}

\vskip 1.7 cm

\begin{center}
{\bf \Large Mapping the $G$-structures and supersymmetric vacua of\\[8pt]
five-dimensional $\mathcal N=4$ supergravity}

\vskip 1cm

{\bf \large James T. Liu,  Manavendra Mahato and Diana Vaman}

\vskip 1cm
{\it Michigan Center for Theoretical Physics,
University of Michigan\\
Ann Arbor, MI 48109--1040}\\
\vskip .5 cm
{\tt jimliu,mmahato,dvaman@umich.edu}

\end{center}

\vspace{1cm}

\begin{abstract}

We classify the supersymmetric vacua of $\mathcal N=4$, $d=5$ supergravity
in terms of $G$-structures. We identify three classes of solutions: with
$\mathbb R^3$, SU(2) and generic SO(4) structure.  Using the Killing spinor
equations, we fully characterize the first two classes and partially solve
the latter.  With the $\mathcal N=4$ graviton multiplet decomposed in terms
of $\mathcal N=2$ multiplets: the graviton, vector and gravitino multiplets,
we obtain new supersymmetric solutions corresponding to turning on fields
in the gravitino multiplet. These vacua are described in terms of an SO(5)
vector sigma-model coupled with gravity, in three or four dimensions.  A
new feature of these $\mathcal N=4$ vacua, which is not seen from an
$\mathcal N=2$ point of view, is the possibility for preserving more exotic
fractions of supersymmetry.  We give a few concrete examples of these new
supersymmetric (albeit singular) solutions.  Additionally, we show how by
truncating the $\mathcal N=4$, $d=5$ set of fields to minimal supergravity
coupled with one vector multiplet we recover the known two-charge solutions.

\end{abstract}

\end{titlepage}

\section{Introduction}

One of the most important principles underlying much of physics is
the use of symmetries as a means of classifying and understanding
physical phenomena.  This is especially true in the theoretical realm,
where the use of standard symmetries such as Lorentz and gauge invariance
has played a key r\^ole in the development of quantum field theories of
the Standard Model and beyond.  Along these lines, the use of supersymmetry
has been at the forefront of many recent explorations into both formal
string theory as well as string and particle phenomenology.  After all,
supersymmetry is a natural extension of the Poincar\'e symmetry of
spacetime, and furthermore may be argued to be a natural ingredient of
any realistic theory of quantum gravity.

Given an underlying supersymmetric theory, it is of course expected that
many interesting vacua or configurations may break some or all of the
supersymmetries.  In fact, it is precisely the BPS states, namely the
configurations with partially broken symmetry, that are of much interest
in the field.  This is because potential corrections to these objects
are much better controlled, whether through multiplet shortening
or related non-renormalization theorems.  As a result, BPS states are
an important tool in the study of strong/weak coupling dualities, where
otherwise one would naturally expect large corrections to appear.

Through the use of duality, BPS objects such as black holes and branes often
have multiple descriptions.  On one side of a duality, they may be constructed
as exact solutions within a particular supergravity framework, while
on the other side they may be fundamental strings, D-branes or other such
objects.  From this point of view, the construction and classification of
exact BPS solutions has certainly given rise to many important advances.
This is especially true in the development of our understanding of D-branes
and the counting of black hole microstates, both of which depended greatly
on the existence of corresponding supergravity solutions.

In addition, the classification of supersymmetric vacua is of current
interest in the program of mapping out the string landscape.  Ideally one
would like to be able to answer the question of what string, brane or
flux compactifications are possible that lead to realistic $\mathcal N=1$
models in four dimensions.  While this has been answered in the conventional
perturbative heterotic picture by SU(3) holonomy ({\it i.e.}~Calabi-Yau)
manifolds, less is known in the presence of fluxes and branes.  Nevertheless,
progress is being made in this direction, and much of that has been due to
better understanding of fluxes and $G$-structures.

Much of the recent work on classifying and constructing supersymmetric
configurations involves the invariant tensor analysis originally
developed in \cite{Tod:1983pm,Tod:1995jf} and further developed in
\cite{Gauntlett:2002sc,Gauntlett:2002nw,Gauntlett:2002fz,Gauntlett:2003wb}.
In this analysis, one first postulates the existence of a Killing spinor
$\epsilon$.  Given such a Killing spinor, one is then able to construct
a set of invariant tensors formed out of spinor bilinears.  The existence
of such invariant tensors ensures the existence of a preferred $G$-structure.
This $G$-structure, along with its intrinsic torsion classes then provides
a framework for the classification of all supersymmetric solutions.  To
proceed to an actual construction, one must examine the `differential
relations' which follow from the actual Killing spinor equations.  Here
we note that solving these relations to arrive at an explicit field
configuration is often the most challenging step in the construction.
Finally, as partially broken supersymmetry does not necessarily imply the
full set of equations of motion, one may have to examine an appropriate
subset of them to complete the construction.  This is generally the origin
of the resulting second order `harmonic function' equations.

The invariant tensor analysis has been particularly fruitful in theories
with eight supercharges.  This includes four-dimensional $\mathcal N=2$
ungauged \cite{Tod:1983pm,Tod:1995jf} and gauged
\cite{Caldarelli:2003pb,Cacciatori:2004rt} supergravity,
five-dimensional $\mathcal N=2$ (minimal) ungauged \cite{Gauntlett:2002nw}
and gauged \cite{Gauntlett:2003fk,Gutowski:2004yv,Gutowski:2005id}
supergravity, and six-dimensional $\mathcal N=(1,0)$ ungauged
\cite{Gutowski:2003rg} and gauged \cite{Cariglia:2004kk} supergravity.
The classification in terms of $G$-structures is also reasonably well
developed for eleven dimensional supergravity, with $32$ supercharges.
However, the actual construction of all possible solutions is a
great technical challenge in the models with more supersymmetries,
as there are many more bosonic degrees of freedom that must be pinned
down.  One way to overcome these difficulties is to impose additional
isometries on the BPS solutions.  This method has been used to great
success in the bubbling AdS work of \cite{Lin:2004nb}.  However, this
still leaves open the question of what is the full class of solutions
without any restriction on the isometries.

Given a well developed set of techniques applied to theories with
eight supercharges, it is then natural to explore the construction
of all supersymmetric solutions in theories with $16$ supercharges as
an intermediate step on the way to theories with $32$ supercharges.
Proceeding towards this goal, a $G$-structure classification for
seven-dimensional supergravity was given in
\cite{Cariglia:2004qi,MacConamhna:2004fb}, and a construction of
all supersymmetric configurations of $\mathcal N=4$ ungauged
supergravity in four dimensions was recently given in
\cite{Bellorin:2005zc}.

In this paper, we continue the study of theories with $16$ supercharges
by constructing all supersymmetric solutions to $\mathcal N=4$ ungauged
supergravity in five dimensions.  This theory is in fact closely
related by dimensional reduction to $\mathcal N=4$ supergravity in
four dimensions, which was investigated in \cite{Bellorin:2005zc}.
However, the present five-dimensional case is somewhat more general,
and we find that solutions break up into three classes, namely those with
either: $i$) a timelike Killing vector and SU(2) structure,
$ii$) a timelike Killing vector and SO(4) structure, or
$iii$) a null Killing vector and $\mathbb R^3$ structure.
All such solutions are characterized by an SO(5) sigma model corresponding
to a vector $u^a$ ($a=1,\ldots,5$) with unit norm.  In the rigid
(constant $u^a$) case, the generic solution preserves $1/4$ of the
supersymmetries, although special configurations preserve either $1/2$ or
all of the supersymmetries.  This rigid case admits a natural $\mathcal N=2$
interpretation in terms of supergravity coupled to a single vector
multiplet.

The non-rigid cases (whether for timelike or null Killing isometries)
are rather more unusual, as they have no direct correspondence in the
$\mathcal N=2$ theory.  From an $\mathcal N=2$ perspective, these cases
correspond to exciting the gauge fields in the gravitino multiplet.  As a
result, they give rise to true $\mathcal N=4$ configurations.  Furthermore, it
appears that these non-rigid solutions may preserve any of $0, 1, 2, 3, 4, 6,
8$ or $16$ of the $\mathcal N=4$ supersymmetries.  We present
some examples, although we have yet to find a completely regular solution
in this class which is free of all singularities.

In the following section, we review the $\mathcal N=4$, $d=5$ ungauged
supergravity theory and proceed to construct the spinor bilinears.  Use
of the Fierz identities allows us to deduce the $G$-structure classification
indicated above.  In Section 3, we specialize to the timelike Killing
vector case and present a complete investigation of the solutions
preserving a SU(2) structure and say a few words about the SO(4) structure
case.  Following this, we take up the null case in Section 4.
Finally, we conclude in Section 5 with a few concrete examples of
non-rigid solutions.  Some of the technical details are
relegated to the appendices.  In particular, Appendix~A contains a set
of important Fierz identities, and Appendix~B tabulates the differential
identities following from the Killing spinor equations.

\section{${\cal N}=4$ supergravity and $G$-structures}

Five-dimensional $\mathcal N=4$ supergravity was first constructed in
\cite{Awada:1985ep}, and is formulated in terms of a five-dimensional
USp(4) symplectic Majorana spinor $\epsilon^i$.  In the minimal ungauged
case, the bosonic fields consist of a metric, a scalar $\phi$, and six
abelian gauge fields $A_\mu^{[ij]|}$ and $B_\mu$ (with field strengths
$F_{\mu\nu}^{[ij]|}$ and $G_{\mu\nu}$), transforming under USp(4)
as the $\mathbf 5$ and $\mathbf 1$, respectively.  The fermionic fields
are comprised by the four gravitini $\psi_\mu^i$ and the four dilatini
$\chi^i$, both of which transform as the $\mathbf 4$ of USp(4).

Up to terms quartic in fermions, the Lagrangian is
\bea
e^{-1}{\cal L}&=&R-\fft12 (\partial_\mu\phi)^2
-\frac 18 e^{\fft2{\sqrt6}\phi}(F^{ij}_{\mu\nu})^2
-\frac 14 e^{-\fft4{\sqrt6}\phi}(G_{\mu\nu})^2
-\frac 12\bar\psi_\mu^i \Gamma^{\mu\nu\rho}\nabla_\nu\psi_i
-\frac 12 \bar \chi^i\Gamma^\mu\nabla_\mu\chi_i\nn\\
&&
+\frac {1}{16}\epsilon^{\mu\nu\rho\sigma\lambda}
F_{\mu\nu}^{ij}F_{\rho\sigma \, ij}B_\lambda -
\frac{i}{2\sqrt 2} e^{-\fft2{\sqrt6}\phi}\bar\chi^i\Gamma^\mu
\Gamma^\nu\psi_{\mu\,i}\partial_\nu\phi
+\frac{1}{4\sqrt 3}e^{\fft1{\sqrt6}\phi}\bar\chi^i\Gamma^\mu
\Gamma^{\rho\sigma}\psi_{\mu}^jF_{\rho\sigma\,ij}\nn\\
&&
-\frac{1}{4\sqrt 3}e^{-\fft2{\sqrt6}\phi}\bar\chi^i\Gamma^\mu
\Gamma^{\rho\sigma}\psi_{\mu\,i}G_{\rho\sigma}
-\frac{i}{24}e^{\fft1{\sqrt6}\phi}\bar\chi^i\Gamma^{\rho\sigma}\chi^j
F_{\rho\sigma\,ij}
+\frac{5}{48}e^{-\fft2{\sqrt6}\phi}\bar\chi^i\Gamma^{\rho\sigma}\chi_i
G_{\rho\sigma}\nn\\
&&
-\frac{i}{8}e^{\fft1{\sqrt6}\phi}\left[\bar\psi_\mu^i\Gamma^{\mu\nu\rho
\sigma}\psi_\nu^j+2\bar\psi^{\rho\,i}\psi^{\sigma\,j}\right]F_{\rho\sigma\,ij}
-\frac{i}{16}e^{-\fft2{\sqrt6}\phi}\left[\bar\psi_\mu^i\Gamma^{\mu\nu\rho
\sigma}\psi_{\nu\,i}+2\bar\psi^{\rho\,i}\psi^{\sigma}_i\right]G_{\rho\sigma},
\nonumber\\
\eea
where we have rescaled some of the fields of \cite{Awada:1985ep}.  Note
that we work with signature $(-,+,+,+,+)$.

The supersymmetry transformations are given by
\bea
&&\delta e_{\mu}^m=\frac 14\bar \epsilon^i\Gamma^m \psi_{\mu\,i},\\
&&\delta \phi=\frac{i}{\sqrt 2}\bar\epsilon^i\chi_i,\\
&&\delta A_\mu^{ij}=-\frac 1{\sqrt 3}\no\left(\bar\epsilon^i\Gamma_\mu\chi^j
+\frac14\Omega^{ij}\bar\epsilon^k\Gamma_\mu\chi_k\right)
-i\no\left(\bar\epsilon^i\psi_\mu^j+\frac14\Omega^{ij}\bar\epsilon^k
\psi_{\mu \,k}\right),\\
&&\delta B_\mu=\frac{1}{2\sqrt 3}\2\bar\epsilon^i\Gamma_\mu\chi_i-\frac i4\2
\bar\epsilon^i\psi_{\mu\,i},\\
&&\delta\psi_{\mu\,i}=\nabla_\mu\epsilon_i+\frac{i}{12}F_{\rho\sigma\,ij}
\1\left(\Gamma_{\mu}{}^{\rho\sigma}-4\delta_\mu^\rho\Gamma^\sigma\right)
\epsilon^j+\frac{i}{24}G_{\rho\sigma}\nt\left(\Gamma_{\mu}{}^{\rho\sigma}
-4\delta_\mu^\rho\Gamma^\sigma\right)\epsilon_i,
\label{eq:deltapsi}\\
&&\delta\chi_i=-\frac{i}{2\sqrt 2}\partial_\mu\phi\Gamma^\mu\epsilon_i+
\frac1{4\sqrt3}\1F_{\rho\sigma\,ij}\Gamma^{\rho\sigma}\epsilon^j-\frac
{1}{4\sqrt 3}\nt G_{\rho\sigma}\Gamma^{\rho\sigma}\epsilon_{i},
\label{eq:deltachi}
\eea
up to three-fermi terms in the gravitino and dilatino variations.  Here
$\Omega_{ij}$ is the real antisymmetric USp(4) invariant tensor satisfying
$\Omega^{ij}\Omega_{jk}=-\delta^i_k$.  In particular,
$\Omega_{ij}$ is used to raise and lower the USp(4) indices according to
the northwest-southeast rule
\be
V^i=\Omega^{ij}V_j,\qquad V_i=V^j\Omega_{ji}.
\ee
All spinors are symplectic Majorana, obeying
\be
\bar\lambda^i=(\lambda^i)^T C,
\ee
where $\bar\lambda^i\equiv(\lambda_i)^*\Gamma_0$, and
$C$ is the real antisymmetric charge conjugation matrix.  It is also
useful at this stage to note the Majorana flip condition
\begin{equation}
\overline\chi^i\Gamma^{\mu_1\cdots\mu_n}\lambda^j=-(-)^{n(n-1)/2}
\overline\lambda^j\Gamma^{\mu_1\cdots\mu_n}\chi^i.
\label{eq:mflip}
\end{equation}

In what follows, we find it convenient to use the isomorphism between
the USp(4) and SO(5) groups to convert the USp(4) valued indices
$i=1,\dots, 4$ to SO(5) ones $a=1,\dots, 5$.  This may be accomplished
by introducing a set of matrices $T^{a\,i}{}_j$ satisfying the
SO(5) Clifford algebra $\{T^a,T^b\}^i{}_j=2\delta^{ab}\delta^i_j$.
This allows us to convert expressions with USp(4) index pairs into
ones involving purely vectorial SO(5) quantities.

\subsection{Construction of the invariant tensors}

Following
\cite{Gauntlett:2002sc,Gauntlett:2002nw,Gauntlett:2002fz,Gauntlett:2003wb},
for a purely bosonic background, we first assume the existence of a
commuting Killing spinor $\epsilon^i$ satisfying the Killing spinor equations
\be
\delta \psi_{\mu \,i}=0,\qquad \delta \chi_i=0,
\ee
where the gravitino and dilatino expressions are given by (\ref{eq:deltapsi})
and (\ref{eq:deltachi}), respectively.  Given such a spinor,
we may construct a complete set of invariant tensors formed out of bilinears
of $\epsilon^i$.  In terms of irreducible USp(4) representations, we
define the following bispinors:
\begin{eqnarray}
f^{[ij]|}+f\Omega^{ij}&=&i\bar \epsilon^i\epsilon^j,\nonumber\\
V_\mu^{[ij]|}+K_\mu\Omega^{ij}&=&\bar\epsilon^i\Gamma_\mu\epsilon^j,\nonumber\\
\Phi^{(ij)}_{\mu\nu}&=&i\bar\epsilon^i\Gamma_{\mu\nu}\epsilon^j.
\label{eq:usp4bispin}
\end{eqnarray}
The factors of $i$ have been inserted so that the bispinors are real-valued
tensors, and the (anti-) symmetry properties follow from the Majorana
flip relation (\ref{eq:mflip}).

We note that the total number of bispinor components is given by counting
the number of elements in the matrix $\epsilon^i_\alpha\epsilon^j_\beta$,
which is symmetric under the interchange of the combined indices $(i,\alpha)$
with $(j,\beta)$.  This comes out to $136$, which equals the number of
components in (\ref{eq:usp4bispin}).  More explicitly, the scalar $f$ and
one-form $K$ are USp(4) singlets, while the scalars $f^{ij}$ and one-forms
$V^{ij}$ transform in the ${\bf 5}$ of USp(4), and the two-form $\Phi^{ij}$
belongs to the ${\bf 10}$ representation.  As indicated above, we prefer
to use a SO(5) notation for the bispinors
\begin{eqnarray}
f^a&=&\fft{i}4T^a_{ij}(\bar\epsilon^i\epsilon^j),\qquad\kern2em
f=-\fft{i}4(\bar\epsilon^i\epsilon_i),\nonumber\\
V_\mu^a&=&\fft14T^a_{ij}(\bar\epsilon^i\Gamma_\mu\epsilon^j),\qquad
K_\mu=-\fft14(\bar\epsilon^i\Gamma_\mu\epsilon_i),\nonumber\\
\Phi_{\mu\nu}^{ab}&=&\fft{i}4T^{ab}_{ij}(\bar\epsilon^i\Gamma_{\mu\nu}
\epsilon^j),
\label{eq:so5bispin}
\end{eqnarray}
in which case the $\mathbf5$ and $\mathbf{10}$ of SO(5) is manifest.
Since the underlying spinor $\epsilon^i$ contains only $16$ real
components, not all $136$ components of the above tensors are independent.
As a result, there are numerous algebraic identities (derived through
Fierzing) relating the above tensors to each other.  An important set of
such algebraic identities is presented in Appendix~A.

\subsection{The $G$-structure classification}

For a pure geometry solution, any time the background admits a Killing
spinor $\epsilon$ satisfying $\nabla_\mu\epsilon=0$, there is an
associated Killing vector of the form $(\bar\epsilon\Gamma^\mu\epsilon)$.
This guarantees the existence of at least one isometry associated with
(partially) unbroken supersymmetry.  In the present case, it is easy to
show that, even in the presence of additional fields, the vector $K^\mu$
defined in (\ref{eq:so5bispin}) remains a Killing vector, as it satisfies
the Killing equation (\ref{eq:killing}).  Furthermore, as demonstrated at
the end of Appendix B, this isometry of the metric extends to the entire
solution.

The preferred Killing vector $K^\mu$ plays a fundamental r\^ole in the
identification of the structure group from the invariant tensors.  To
proceed, we note that the norm of $K^\mu$ is easily obtained through
the Fierz identities.  In particular, expression (\ref{eq:KKf}),
namely
\be
K^\mu K_\mu=-(f^a f^a),
\ee
demonstrates that the Killing vector is either timelike or null (as
expected for supersymmetric backgrounds).  The classification thus
splits into two cases.

\subsubsection{The timelike case}
\label{sec:timelike}

For the timelike case, we take $|K|^2=-(f^a)^2<0$.  In this case, the
SO(5) vector defined by $f^a$ is non-vanishing, and may be used to
parameterize the breaking of SO(5) into SO(4).  We make this explicit by defining the projection
\begin{equation}
\Pi_4^{ab}=\delta^{ab}-u^au^b,
\label{eq:so4proj}
\end{equation}
where $u^a$ is the normalized SO(5) vector $u^a=f^a/|f^a|$.  This
projection allows us to decompose the one-forms $V^a$ under
$\mathbf5\to\mathbf4+\mathbf1$ as
\begin{equation}
V_\mu^a=u^aV_\mu^{(1)}+V_\mu^{(4)\,a},
\end{equation}
where $V_\mu^{(1)}=u^aV_\mu^a$ and $V_\mu^{(4)\,a}=\Pi_4^{ab}V_\mu^b$.
Use of the Fierz identity
(\ref{eq:fKf}), namely $f K_\mu=f^a V_\mu^a$, then shows that $V_\mu^{(1)}$
is aligned with $K_\mu$.  This gives
\begin{equation}
V_\mu^a=\fft{ff^a}{(f^b)^2}K_\mu+V_\mu^{(4)\,a}.
\label{eq:vdecomp}
\end{equation}
Furthermore, projecting (\ref{eq:KVf}) onto SO(4) demonstrates that
$i_KV^{(4)\,a}=0$.  This indicates that the above decomposition of
$V_\mu^a$ is onto the timelike direction specified by $K^\mu$ and its
orthogonal spacelike hyperplanes.

The SO(4) valued one-forms $V_\mu^{(4)\,a}$ furthermore satisfy the
conditions
\begin{eqnarray}
V_\mu^{(4)\,a}V^{(4)\,\mu\,b}&=&\left((f^c)^2-f^2\right)\Pi_4^{ab},\nonumber\\
V_\mu^{(4)\,a}V_\nu^{(4)\,a}&=&\left((f^c)^2-f^2\right)
\left(g_{\mu\nu}-K_\mu K_\nu/|K|^2\right),
\label{eq:vierbeins}
\end{eqnarray}
which arise from the Fierz identities (\ref{eq:VVf}) and (\ref{eq:vmuavnua}).
Note that, by taking one additional contraction of either equation, we see
that $\bigl|V_\mu^{(4)\,a}\bigr|^2=4\left((f^c)^2-f^2\right)$.  Since
$V_\mu^{(4)\,a}$ are everywhere spacelike (or vanishing), we are led to
deduce that
\begin{equation}
(f^a)^2\ge f^2.
\label{eq:ffacond}
\end{equation}
Furthermore, the $V_\mu^{(4)\,a}$ must vanish identically when this
inequality is saturated.

If desired, we may make an explicit choice of coordinates so that
$K^\mu\partial_\mu=\partial_t$.  This allows us to express the
five-dimensional metric as a foliation of four-dimensional hypersurfaces
\be
ds^2=-(f^a)^2 (dt+\omega)^2+\fft1{\sqrt{(f^a)^2}} h_{mn} dx^m dx^n,
\label{eq:5dtime}
\ee
In this case, the one-form associated with $K_\mu$ is simply
$K_\mu dx^\mu=-(f^a)^2(dt+\omega)$.  However, we note that the following
discussion is completely general, and does not depend on any particular choice
of coordinates.

Turning next to the two-form $\Phi^{ab}$, we again use $u^a$ to project
it onto invariant SO(4) components
\begin{equation}
\Phi^{ab}=u^a\Phi^{(4)\,b}-u^b\Phi^{(4)\,a}+\Phi^{(6)\,ab},
\end{equation}
under $\mathbf{10}\to\mathbf4+\mathbf6$.  As above, the $\mathbf4$ and
$\mathbf6$ can be disentangled by projecting with combinations of $\Pi_4^{ab}$.
Combining this SO(5) decomposition with the Fierz identity (\ref{eq:kphi}),
we find that
\begin{equation}
\Phi^{ab}=\fft{f^a}{(f^c)^2}K\wedge V^{(4)\,b}-\fft{f^b}{(f^c)^2}
K\wedge V^{(4)\,a} + \Phi^{(6)\,ab}.
\end{equation}
The components valued in the $\mathbf6$ of SO(4) satisfies
$i_K\Phi^{(6)\,ab}=0$, and hence live on surfaces orthogonal to $K^\mu$.
In fact, contraction of (\ref{eq:kphi}) with $K^\mu$ yields a condition
\begin{equation}
(f^c)^2\Phi^{(6)\,ab}-\ft12\epsilon^{abcde}ff^c\Phi^{(6)\,de}
=*(K\wedge V^{(4)\,a}\wedge V^{(4)\,b}),
\end{equation}
while the identity (\ref{eq:VwedgeVf}) gives directly
\begin{equation}
f\Phi^{(6)\,ab}+*(K\wedge\Phi^{(6)\,ab})=0.
\label{eq:phiasd}
\end{equation}
Additional use of (\ref{eq:kphi}) then leads to the expressions
\begin{eqnarray}
\left((f^c)^2-f^2\right)\Phi^{(6)\,ab}&=&-fV^{(4)\,a}\wedge V^{(4)\,b}
+*(K\wedge V^{(4)\,a}\wedge V^{(4)\,b}),\nonumber\\
\left((f^c)^2-f^2\right)\ft12\epsilon^{abcde}f^c\Phi^{(6)\,de}&=&
-(f^c)^2V^{(4)\,a}\wedge V^{(4)\,b}+f*(K\wedge V^{(4)\,a}\wedge V^{(4)\,b}).
\label{eq:phisolv}
\end{eqnarray}
Combining the two above equations demonstrates that the two-form
$V^{(4)\,a}\wedge V^{(4)\,b}$ must satisfy the joint spacetime and internal
SO(4) anti-self-duality condition
\begin{equation}
\left((f^c)^2-f^2\right)
\left(\delta_\mu^{[\rho}\delta_\nu^{\sigma]}\delta^{ac}\delta^{bd}
+(\ft12\epsilon^{abcde}u^e)(\ft12\epsilon_{\mu\nu}{}^{\rho\sigma\lambda}
K_\lambda)\right)V_\rho^{(4)\,c}V_\sigma^{(4)\,d}=0.
\end{equation}

{\it SU(2) structure.}
Until now, we have not placed any further restrictions on $f$ and $f^a$
other than the inequality (\ref{eq:ffacond}).  It ought to be clear from
the above, however, that we ought to distinguish between two subcases of
the general timelike case, depending on whether the inequality is
saturated or not.  Consider first the case
\be
(f^a)^2=f^2.
\label{su2strcond}
\ee
In this case, $V^{(4)\,a}$ vanishes, and we are left with
\begin{equation}
V^a=u^aK\qquad\hbox{and}\qquad \Phi^{ab}=\Phi^{(6)\,ab},
\end{equation}
where
\begin{equation}
i_K\Phi^{(6)\,ab}=0,\qquad
\Phi^{(6)\,ab}=\ft12\epsilon^{abcde}u^c\Phi^{(6)\,de}.
\end{equation}
This indicates that the one-forms $V^a$ are aligned with $K$, and that the
two-forms $\Phi^{ab}$ are both transverse to $K^\mu$ and take values in
the self-dual $\rm SU(2)_+$ inside SO(4).  The hyper-K\"ahler structure
may be obtained from the Fierz identity (\ref{eq:hks}), specialized to
the present case:
\begin{equation}
\Phi_{mn}^{ab}\Phi^{n\,cd}_{\;\;p}=
-h_{mp}\bigl[\Pi_4^{ac}\Pi_4^{bd}-\Pi_4^{ad}\Pi_4^{bc}+\epsilon^{abcde}u^e\bigr]
+\bigl[\delta^a_e\delta^c_f\Pi_4^{bd}+\delta^b_e\delta^d_f\Pi_4^{ac}
-\delta^a_e\delta^d_f\Pi_4^{bc}-\delta^b_e\delta^c_f\Pi_4^{ad}\bigr]
\Phi_{mn}^{ef},
\end{equation}
where space indices are raised with the four-dimensional metric $h_{mn}$
of (\ref{eq:5dtime}).  Moreover, from (\ref{eq:phiasd}),
we see that the two-forms are anti-self-dual on the four-dimensional base.
This presents a curious connection between the spacetime and internal
indices of $\Phi_{\mu\nu}^{(6)\,ab}$, as it resides in both the tangent
space group $\rm SU(2)_-\subset SO(4)\subset SO(4,1)$ and the internal
symmetry group $\rm SU(2)_+\subset SO(4)\subset SO(5)$.

This case is nearly identical to the corresponding one for timelike
configurations of minimal $\mathcal N=2$ supergravity \cite{Gauntlett:2002nw}.
The combination of a timelike Killing vector $K^\mu$ along with an
$SU(2)_+$ triplet of two-forms $\Phi^{(6)\,ab}$ with $i_K\Phi^{(6)\,ab}=0$
guarantees the existence of a preferred SU(2) structure.
However, the distinction between the ${\mathcal N}=2$ theory with
$\rm USp(2)\simeq SU(2)$ and the ${\mathcal N}=4$ theory with
$\rm USp(4)\simeq SO(5)$ is clear: in order to identify an SU(2) structure
in the $\mathcal N=4$ case, we had to impose the additional constraint
(\ref{su2strcond}) on the bilinears.  Furthermore, as we will see in
Section 4, unlike for the $\mathcal N=2$ case, where the base had actually
SU(2) holonomy, here we find only the weaker condition of SU(2) structure.

{\it SO(4) structure.} Finally, we note that if $(f^a)^2>f^2$, then
the nature of the solution is strikingly different.  In particular, from
(\ref{eq:vierbeins}), we see that the SO(4) valued one-forms $V^{(4)\,a}$
serve as vielbeins on the four-dimensional base transverse to $K^\mu$.
More precisely, by defining
\begin{equation}
e^a=\fft1{\sqrt{\left((f^b)^2-f^2\right)}}V^{(4)\,a}
\label{eq:viers}
\end{equation}
we are led to a natural five-dimensional metric (\ref{eq:5dtime}) of the
form $ds^2=-(e^0)^2+(e^a)^2$ with $e^0=\sqrt{(f^a)^2}(dt+\omega)$ and
$e^a$ given above.
So long as $(f^a)^2>f^2$, the two-forms $\Phi^{(6)\,ab}$ are completely
determined by
\begin{equation}
\Phi^{(6)\,ab}=-fe^a\wedge e^b-|f^a|*(e^0\wedge e^a\wedge e^b),
\end{equation}
where we have taken $K=-(f^a)^2(dt+\omega)=-|f^a|e^0$.
Note that these two-forms do not have any particular (anti-)self-duality
properties on the base, as $|f^a|\ne |f|$.  

The timelike Killing vector $K^\mu$, in conjunction with the four-dimensional
base metric $ds_4^2=(e^a)^2$, indicates that the generic SO(5) structure
is reduced to SO(4).  Furthermore, this SO(4) structure case has no
counterpart in the $\mathcal N=2$ analysis.  As a result, supersymmetric
configurations falling in this class presumably would not admit a purely
$\mathcal N=2$ interpretation.  Of course, a more detailed examination
would be in order to see if this is really the case.

\subsubsection{The null case}
\label{sec:null}

The null case is given by $|K|^2=-(f^a)^2=0$.  From this, we infer
that the five scalars $f^a$ are all vanishing.  Additionally, from the
identity (\ref{eq:fKf}), namely $fK_\mu=f^a V_\mu^a$, we find that $f$
vanishes as well (since we assume $K_\mu$ to be everywhere non-vanishing).
Next, we may use the identity (\ref{eq:faphiab}), given in form notation
as $K\wedge V^a=-\Phi^{ab}f^b$ ($=0$ when $f^b=0$),
to demonstrate that $V^a$ is aligned with $K$.  This allows us to write
\begin{equation}
V_\mu^a=u^aK_\mu.
\end{equation}
The norm of the SO(5) vector $u^a$ is determined from (\ref{eq:vmuavnua})
\be
V^a_\mu V^a_\nu=K_\mu K_\nu + g_{\mu\nu}((f^a)^2-f^2)
\ee
to be equal to one:
\be
u^a u^a=1.
\ee
Therefore (just as in the timelike case) the $u^a$'s parametrize an
SO(5)/SO(4) coset.

Tuning now to the two-forms $\Phi^{ab}$, we recall that they generically
transform as the $\mathbf{10}$ of SO(5).  However (\ref{eq:vmuaphi})
\be
V_\mu^a \Phi^{ab}_{\nu\lambda}=-2g_{\mu[\nu}(fV_{\lambda}^b-f^bK_{\lambda]})
+\epsilon_{\mu\nu\lambda}{}^{\rho\sigma}V_\rho^bK_\sigma^{\vphantom{b}}
\ee
reveals that the number of independent two-forms $\Phi^{ab}$ is in fact
constrained by
\be
u^a \Phi^{ab}=0,
\ee
and so we are left with only six two-forms, corresponding
to keeping the $\mathbf6$ in the decomposition $\mathbf{10}\to\mathbf6
+\mathbf4$ under $\rm SO(5)\to SO(4)$.  In fact, these six components
are not all independent, as can be seen by consideration of the
additional Fierz identity (\ref{eq:kphi})
\be
K_\mu\Phi_{\nu\lambda}^{ab}=-2g_{\mu[\nu}(f^a V_{\lambda}^b-f^b
V_{\lambda]}^a)-\epsilon_{\mu\nu\lambda}{}^{\rho\sigma}V_{\rho}^aV_\sigma^b
+\ft12\epsilon^{abcde}V_\mu^c \Phi_{\nu\lambda}^{de}.
\ee
This further constrains $\Phi^{ab}$ to satisfy a self-duality condition
in group space
\be
\Phi^{ab}=\frac12 \epsilon^{abcde} u^c \Phi^{de}.
\ee
And so we are left with the $\mathbf3_+$ under the complete decomposition
$\mathbf{10}\to\mathbf6+\mathbf4\to(\mathbf3,\mathbf1)+(\mathbf1,\mathbf3)
+(\mathbf2,\mathbf2)$ of $\rm SO(5)\to SO(4)\to SU(2)_+\times SU(2)_-$.
(Note that this $\rm SU(2)_+$ is an internal symmetry group, and is at least
superficially unrelated to the structure of the space.)

Finally, using (\ref{eq:kmuphimu}) and (\ref{eq:epskphi})
\bea
K^\mu\Phi_{\mu\nu}^{ab}&=&-2f^{[a}V_\nu^{b]},\nonumber\\
\epsilon_{\mu\nu\rho\sigma\lambda} K^{\rho}\Phi^{\sigma\lambda\, ab}&=&
-\epsilon^{abcde} f^c\Phi_{\mu\nu}^{de},
\eea
we conclude that
\begin{equation}
i_K\Phi^{ab}=i_K*\Phi^{ab}=0.
\label{eq:ikik*phi}
\end{equation}
This combination of a null Killing vector $K^\mu$ along with three
independent two-forms $\Phi^{ab}$ satisfying (\ref{eq:ikik*phi}) demonstrates
that there is an $\mathbb R^3$ structure associated with this null Killing
vector case.

Although the above is sufficient to demonstrate $\mathbb R^3$ structure,
we find it useful to make more explicit choices for the purpose of
constructing solutions, which is taken up in Section~\ref{sec:nullsoln}.
In particular,
using the appropriate differential identities below, we may demonstrate
that $K$ is hypersurface orthogonal, and that it can be chosen to be
$K^\mu\partial_\mu=\partial_v$.  This allows us to
write the five-dimensional metric as
\be
ds^2 = H^{-1}({\cal F} du^2+2\,du\,dv)-H^2h_{mn}(dy^m+a^m du)(dy^n+a^n du).
\ee
We can now use the fact that the two-forms $\Phi^{ab}$ are aligned
with $K$ to introduce a set of one-forms $X^{ab}$ according to
\be
\Phi^{ab}=K\wedge X^{ab}.
\ee
These three independent one-forms $X^{ab}$ reside on the three-dimensional
base (with metric $h_{mn}$), and satisfy the multiplication rule
\be
X_m^{ac} X_n^{bc}=-\epsilon_{mnp}X^{p\, ab}+\Pi_4^{ab}h_{mn},
\label{eq:x multiplication}
\ee
which may be obtained from (\ref{eq:42fierz}).  Note that $\Pi_4^{ab}$ is
given by (\ref{eq:so4proj}), and is a projection onto
$\rm SO(4) \subset SO(5)$.  Thus we are led to the conclusion that, in the null
Killing vector case, the five-dimensional metric can be written in terms
of a three-dimensional base (hypersurfaces orthogonal to the Killing vector),
with three independent one-forms $X^{ab}$ satisfying SO(4) self-duality
with respect to $u^a$ residing on it.

Note that it is important to keep in mind that $\mathbb R^3$ structure does
not guarantee the closure of these one-forms $X^{ab}$.  Of course, in the
minimal $\mathcal N=2$ theory, the corresponding one-forms constructed
from the original two-forms through similar steps as above were found
to be closed \cite{Gauntlett:2002nw}.  In the $\mathcal N=4$ case,
however, this will happen only in special circumstances; this issue will be
addressed at length in Section~\ref{sec:nullsoln}, when we take up the
differential identities in the null case.

\subsection{Differential identities}

Until now, we have mainly focused on the algebraic identities and the
resulting structure equations.  As is well known, the existence of an
appropriate set of invariant tensors is sufficient to demonstrate the
appropriate $G$-structure of the system, and this is what we have
accomplished above using the algebraic identities.  Integrability of
these structures, however, falls into the realm of the differential
identities, which we turn to next.

The differential identities encode the content of the Killing spinor
equations, and hence depend explicitly on the model under investigation.
For us, this is the minimal $\mathcal N=4$ supergravity with Killing
spinor equations corresponding to the vanishing of (\ref{eq:deltapsi}) and
(\ref{eq:deltachi}).  However, we anticipate that this analysis could
easily be extended to include the coupling to $\mathcal N=4$ Maxwell
multiplets as well.  These Killing spinor equations may be converted
into differential identities on the bispinors either by multiplication on
the left with $\bar \epsilon \Gamma^{\mu_1\dots\mu_n}$ or by taking the
Hermitian conjugate and then multiplying on the right with
$\Gamma^{\mu_1\dots\mu_n}\epsilon$.  As a result, these equations are
(at most) first order and linear in the bispinors.

Note that, unlike in the case of minimal supergravity, where
there was only the gravitino variation, here we also have the dilatino
equation (\ref{eq:deltachi}) to consider.  The identities resulting
from this dilatino condition are not truly differential, as they are
only algebraic in the bilinears.  We nevertheless denote all such
expressions as `differential identities' to distinguish them from the
algebraic structure equations related to the Fierz identities.  This
notation of differential identities also fits a Kaluza-Klein interpretation,
where the dilatino may be viewed as internal components of the higher
dimensional gravitino.  The loss of the derivative acting on the
dilatino is then attributed to the zero-mode nature of the higher dimensional
gravitino living on the compactification manifold.

The complete set of differential identities are tabulated in Appendix B.
This will provide a basis of the analysis in the following section
for the timelike Killing vector case and Section~\ref{sec:nullsoln} for
the null Killing vector case.

\section{The timelike case}
\label{sec:timesoln}

As indicated above, the timelike case falls into two categories, depending
on the structure being either SU(2) or SO(4).  We focus mainly on the SU(2)
structure case, but will say a few words about the SO(4) structure solutions
at the end of this section.

\subsection{Timelike with SU(2) structure}
\label{sec:timesolnsu2}

The SU(2) structure case
arises when $f^2=(f^a)^2$, and is the most direct generalization of the
analogous $\mathcal N=2$ situation.  To arrive at the complete solution,
we start with the five-dimensional metric of the form (\ref{eq:5dtime})
\begin{equation}
ds^2=-f^2(dt+\omega)^2+f^{-1}h_{mn}dx^mdx^n,
\end{equation}
where $f$, $\omega=\omega_mdx^m$, and $h_{mn}$ are independent of time
$t$.  This metric admits a natural f\"unfbein basis
\begin{equation}
e^0=f(dt+\omega),\qquad e^{\overline m}=f^{-1/2}\hat e^{\overline m}_m
dx^m,
\end{equation}
where $h_{mn}=\hat e_{m\vphantom{\overline m}}^{\overline m}
\hat e_{n\,\overline m}^{\vphantom{m}}$.  We also note that, with our metric
signature, we have $K^\mu\partial_\mu=\partial_t$ and $K_\mu dx^\mu=-fe^0$.

We now proceed to derive the gauge field strengths $F^a$ and $G$.  To
do so, we start with the decomposition
\begin{equation}
G=\alpha\wedge K+\overline G,\qquad
F^a=\alpha^a\wedge K+\overline F^a,
\end{equation}
where $\alpha$ and $\alpha^a$ are one-forms, and $\overline G$ and
$\overline F^a$ are two-forms on the four-dimensional base satisfying
$i_K\overline G=i_K\overline F^a=0$.  Contractions with the Killing vector
then gives
\begin{eqnarray}
&&i_KG=f^2\alpha,\qquad\kern1em i_K*G=f*_4\overline G,\nonumber\\
&&i_KF^a=f^2\alpha^a,\qquad i_K*F^a=f*_4\overline F^a,
\end{eqnarray}
where $*_4$ is defined with respect to the metric $h_{mn}$ on the base.

The one-forms $\alpha$ and $\alpha^a$ may be obtained in terms of the
scalar quantities
\begin{equation}
\mathcal H_1^{-1}\equiv e^{\fft2{\sqrt6}\phi}f,\qquad
\mathcal H_2^{-1}\equiv e^{-\fft1{\sqrt6}\phi}f,
\label{eq:h1h2def}
\end{equation}
through the use of (\ref{KG}) and (\ref{KF}), respectively.  The result
is
\begin{equation}
\alpha=f^{-2}d(\mathcal H_1^{-1}),\qquad
\alpha^a=-f^{-2}d(u^a\mathcal H_2^{-1}).
\end{equation}
The `magnetic' components $\overline G$ and $\overline F^a$ are somewhat
harder to disentangle.  Nevertheless, use of the two-form differential
identities (\ref{K*G}) and (\ref{K*F}) allows us to solve for the
(four-dimensional) self-dual and anti-self-dual components
\begin{eqnarray}
\mathcal H_1\overline G^-&=&-\mathcal F^-,\nonumber\\
\mathcal H_2\overline F^{a\,-}&=&u^a\mathcal F^-\nonumber\\
\mathcal H_1\overline G^+-2u^a\mathcal H_2
\overline F^{a\,+}&=&-\mathcal F^+,\nonumber\\
\Pi_4^{ab}\overline F^{b\,+}&=&0.
\label{eq:ofg}
\end{eqnarray}
Here $\mathcal F=d\omega$, and $\Pi_4^{ab}=\delta^{ab}-u^au^b$ is the
projection onto SO(4).  One further restriction on $\overline F^{a\,-}$
may be obtained from the identity (\ref{eq:Vphi}), which gives the additional
condition $\Phi_{\mu\nu}^{ab}F^{\mu\nu\,b}=0$.  Noting that the SU(2)
structure along with the Fierz identities ensure that the two-form
$\Phi_{\mu\nu}^{ab}$ is anti-self-dual on the base, we may deduce that
the anti-self-dual component of $\overline F^{a}$ must vanish when
projected with $\Phi_{\mu\nu}^{ab}$.  This gives simply $\Pi_4^{ab}
\overline F^{b\,-}=0$, and when combined with (\ref{eq:ofg}), we see that
$\overline F^{a}$ points only along the $u^a$ direction.

The above relations, (\ref{eq:ofg}) along with the condition $\overline F^a
\equiv u^a\overline F$, allow us to write the gauge field strengths in
terms of $\mathcal F=d\omega$ along with an undetermined self-dual two-form
$\overline F^+$.  In particular, we may see that
\begin{eqnarray}
G&=&-d[\mathcal H_1^{-1}(dt+\omega)]+2\fft{\mathcal H_2}{\mathcal H_1}
\overline F^+,\nonumber\\
F^a&=&d[u^a\mathcal H_2^{-1}(dt+\omega)]+u^a\left(\overline F^+
-\mathcal H_2^{-1}\mathcal F^+\right).
\label{eq:fgsol}
\end{eqnarray}
The Bianchi identities $dG=dF^a=0$ immediately give
\begin{equation}
d\left(\fft{\mathcal H_2}{\mathcal H_1}\overline F^+\right)=0,\qquad
d\left(u^a\left(\overline F^+-\mathcal H_2^{-1}\mathcal F^+\right)\right)=0.
\label{eq:su2bianchi}
\end{equation}
At the same time, the two form equations of motion
\begin{equation}
d(e^{\fft2{\sqrt6}\phi}*F^a)=F^a\wedge G,\qquad
d(e^{-\fft4{\sqrt6}\phi}*G)=F^a\wedge F^a,
\label{eq:2eom}
\end{equation}
yield the conditions
\begin{eqnarray}
\Box_4\mathcal H_1&=&\fft12\left(\overline F^+-\mathcal H_2^{-1}
\mathcal F^+\right)^2,\nonumber\\
\Box_4\mathcal H_2&=&-\fft{\mathcal H_2}{\mathcal H_1}\overline F^+
\left(\overline F^+-\mathcal H_2^{-1}\mathcal F^+\right)
+\mathcal H_2u^a\Box_4u^a,
\label{eq:su2eom}
\end{eqnarray}
along with the SO(5) sigma model equation of motion
\begin{equation}
\Box_4 u^a=u^au^b\Box_4 u^b,
\label{eq:so4sigma}
\end{equation}
on the unit-norm SO(5) vector $u^a$.  Note that $\Box_4$ is the scalar
Laplacian with respect to the metric $h_{mn}$ on the base.

Until now, we have not paid much attention to the conditions on $u^a$.
In addition to the sigma model equation of motion given above, $u^a$
must also satisfy a first order condition
\begin{equation}
\partial_m u^a = \Phi_{mn}^{ab}h^{np}\partial_pu^b.
\label{eq:ucond}
\end{equation}
This condition, along with the expressions for $F^a$ and $G$ given in
(\ref{eq:fgsol}), guarantees that all one-form and two-form differential
identities (\ref{KG}) through (\ref{V*G}) are satisfied.

Finally, the remaining differential identities, and in particular
(\ref{eq:dPhi}) for $\nabla_\mu\Phi_{\nu\lambda}^{ab}$, demand that
\begin{equation}
\hat\nabla_m\Phi_{np}^{ab}=\fft43[\Phi_{m[p}^{bc}u^a\partial_{n]}u^c
+\Phi_{np}^{bc}u^a\partial_mu^c+\fft12\epsilon_{mnp}{}^qu^a\partial_qu^b],
\label{eq:gradphi}
\end{equation}
where all quantities are defined in terms of the metric $h_{mn}$.
Making use of (\ref{eq:ucond}), along with anti-self-duality of
$\Phi_{mn}^{ab}$ and the projection $u^a\Phi_{mn}^{ab}=0$, the above
expression can be written in the form
\begin{equation}
D_m^{\vphantom{a}}\Phi_{pq}^{ab}\equiv\hat\nabla_m^{\vphantom{a}}
\Phi_{pq}^{ab}+\mathcal A_m^{ac}\Phi_{pq}^{cb}
+\mathcal A_m^{bc}\Phi_{pq}^{ac}=0,
\label{eq:compcovar}
\end{equation}
where $\mathcal A_m^{ab}$ is the composite SO(5) connection
\begin{equation}
\mathcal A^{ab}=2u^{[a}du^{b]}.
\label{eq:composite}
\end{equation}
This clearly shows that in the rigid case (where $u^a$ is constant, so
that $\mathcal A_m^{ab}$ vanishes), we have
$\hat\nabla_m^{\vphantom{a}}\Phi_{pq}^{ab}=0$, which implies that the
four-dimensional base has SU(2) holonomy.  However, in the general
non-rigid case, the base only has SU(2) structure.  Note, also, that
the fully antisymmetrized components of (\ref{eq:gradphi}) may be
written as
\begin{equation}
d\Phi^{ab}=*_4\mathcal A^{ab},
\end{equation}
which demonstrates that the composite connection may also be given in
terms of the two-form $\Phi^{ab}$.

Integrability of the covariant derivative $D_m$ gives rise to
\begin{equation}
\hat R_{mnp}{}^r\Phi_{rq}^{ab}+\hat R_{mnq}{}^r\Phi_{pr}^{ab}
+\mathcal F_{mn}^{ac}\Phi_{pq}^{cb}+\mathcal F_{mn}^{bc}\Phi_{pq}^{ac}=0,
\label{eq:timintr}
\end{equation}
where the composite field strength is given by
\begin{equation}
\mathcal F_{mn}^{ab}\equiv2\partial_{[m}\mathcal A_{n]}^{ab}
+2\mathcal A_{[m}^{ac}\mathcal A_{n]}^{cb}
=\partial_mu^a\partial_nu^b-\partial_mu^b\partial_nu^a.
\end{equation}
Contracting (\ref{eq:timintr}) with $\Phi_{mq}^{ab}$ and using the structure
equations
\begin{eqnarray}
\Phi_{mn}^{ab}\Phi_{pq}^{ab}&=&4\epsilon_{mnpq}+4(h_{mp}h_{nq}-h_{mq}h_{np}),
\nonumber\\
\Phi_{mn}^{ab}\Phi^{n\,bc}_{\;\;p}&=&3\Pi_4^{ac}h_{mp}-2\Phi_{mp}^{ac},
\end{eqnarray}
results in the integrability condition $\hat R_{mn}=-\fft12\mathcal F_{mp}^{ab}
\Phi_n{}^{p\,ab}$.  Using (\ref{eq:ucond}), this gives the Einstein equation
\begin{equation}
\hat R_{mn}=-\partial_mu^a\partial_nu^a,
\label{eq:wseins}
\end{equation}
on the base.  The combination of the equation of motion (\ref{eq:so4sigma})
along with the Einstein equation is suggestive of an SO(5) sigma model
coupled to gravity.  However, in the present situation, the gravitational
coupling in (\ref{eq:wseins}) is of the `wrong sign', corresponding to
a negative stress-energy tensor.

The above Einstein equation indicates that in the non-rigid case the
four-dimensional base can no longer be Ricci-flat.  This is an explicit
demonstration that such solutions only have SU(2) structure, and not
holonomy.  Before proceeding, we note that taking the trace of the
Einstein equation (\ref{eq:wseins}) and making use of the fact that
$u^a$ has unit norm (so that $u^a\partial_m u^a=0$) gives us a simple
expression for the four-dimensional curvature scalar
\begin{equation}
\hat R=u^a\Box_4u^a.
\end{equation}
If desired, this allows us to rewrite the scalar equations (\ref{eq:su2eom})
and (\ref{eq:so4sigma}) as
\begin{equation}
\Box_4\mathcal H_1=\ft12(G_2^+)^2,\qquad
(\Box_4-\hat R)\mathcal H_2=\ft12G_1^+G_2^+,\qquad
(\Box_4-\hat R)u^a=0,
\label{eq:scaeom}
\end{equation}
where we have defined the self-dual two-forms $G_1^+$ and
$G_2^+$ by
\begin{equation}
G_1^+=-2(\mathcal H_2/\mathcal H_1)\overline F^+,\qquad
G_2^+=\overline F^+-\mathcal H_2^{-1}\mathcal F^+.
\end{equation}
This demonstrates that $\mathcal H_1$ behaves as a minimally
coupled scalar, while $\mathcal H_2$ behaves as a non-minimally
coupled scalar on the four-dimensional base.

Of course, in addition to the second order equations
(\ref{eq:so4sigma}) and (\ref{eq:wseins}), supersymmetry demands the
stronger first order condition (\ref{eq:ucond}) as well.  As this
condition is somewhat awkward to work with directly (since the two-form
$\Phi^{ab}$ is incompletely specified), it is instructive to directly
examine the Killing spinor equations (\ref{eq:deltapsi}) and
(\ref{eq:deltachi}) for this SU(2) structure timelike solution.
Substituting in the expressions for the gauge fields (\ref{eq:fgsol}),
as well as the definitions for the scalars (\ref{eq:h1h2def}), we obtain
the slightly cumbersome expressions
\begin{eqnarray}
\sqrt{3}f^{-1/2}\delta\chi_i\!\!&=&\!\!
\bigl[i\delta_i^j\hat\gamma^m\partial_m\log\mathcal H_1-\ft14f^{3/2}
\mathcal H_2((2\overline F_{mn}^+-\mathcal H_2^{-1}\mathcal F_{mn}^+)
\delta_i^j+\overline F_{mn}^+u^aT^a_i{}^j)\hat\gamma^{mn}\bigr]P_1\epsilon_j
\nonumber\\
&&-\bigl[i\hat\gamma^m\partial_m\log\mathcal H_2\bigr]P_2\epsilon_i
+\bigl[\ft12f^{3/2}\mathcal F_{mn}^-\hat\gamma^{mn}\bigr]P_3\epsilon_i
+\ft12\Gamma^0(\hat\gamma^m\partial_mu^aT^a_i{}^j)\epsilon_j,\nonumber\\
f^{-1/2}\delta\psi_{t\,i}\!\!&=&\!\!\bigl[-\ft13\delta_i^j\Gamma_0\hat\gamma^m
\partial_m\log\mathcal H_1+\ft1{12}f^{3/2}\mathcal H_2
((\overline F_{mn}^+-2\mathcal H_2^{-1}\mathcal F_{mn}^+)\delta_i^j
+\overline F_{mn}^+u^aT^a_i{}^j)\hat\gamma^{mn}\bigr]P_1\epsilon_j
\nonumber\\
&&-\bigl[\ft23\Gamma_0\hat\gamma^m\partial_m\log\mathcal H_2\bigr]P_2\epsilon_i
-\bigl[\ft16f^{3/2}\mathcal F_{mn}^-\hat\gamma^{mn}\bigr]P_3\epsilon_i
-\ft{i}3(\hat\gamma^m\partial_mu^aT^a_i{}^j)\epsilon_j,\nonumber\\
\delta\psi_{m\,i}-\omega_m\delta\psi_{t\,i}\!\!&=&\!\!
(\mathcal H_1\mathcal H_2^2)^{-1/6}
\bigl[\hat\nabla_m\delta_i^j-\ft12\partial_mu^aT^a_i{}^j\bigr]
(\mathcal H_1\mathcal H_2^2)^{1/6}\epsilon_j\nonumber\\
&&+\bigl[\ft16\delta_i^j(\hat\gamma_m{}^n-2\delta_m^n)\partial_n
\log\mathcal H_1+\partial_mu^aT^a_i{}^j-\ft13f^{3/2}(\mathcal F_{mn}^+
+3\mathcal F_{mn}^-)\delta_i^j\hat\gamma^n\Gamma^0\bigr]P_1\epsilon_j\nonumber\\
&&+\bigl[\ft13\delta_i^j(\hat\gamma_m{}^n-2\delta_m^n)\partial_n
\log\mathcal H_2\bigr]P_2\epsilon_i\nonumber\\
&&+\bigl[\ft13f^{3/2}\mathcal H_2
(-\overline F_{mn}^+(1+2i\Gamma^0)+\mathcal H_2^{-1}\mathcal F_{mn}^-
(1-2i\Gamma^0))\hat\gamma^n\Gamma^0\bigr]P_3\epsilon_i\nonumber\\
&&-\ft{i}6\Gamma^0\hat\gamma_m(\hat\gamma^n\partial_nu^aT^a_i{}^j)\epsilon_j.
\label{eq:timekses}
\end{eqnarray}
Here we have defined the projections
\begin{equation}
P_1=\ft12(1+i\Gamma^0),\qquad P_{2\,i}{}^j=\ft12(\delta_i^j+i\Gamma^0
u^aT^a_i{}^j),\qquad P_{3\,i}{}^j=\ft12(\delta_i^j-u^aT^a_i{}^j).
\label{eq:projs}
\end{equation}
Note, also, that the Dirac matrices $\hat\gamma^m$ are defined with
respect to the base metric $h_{mn}$.

The three projections in (\ref{eq:projs}) are mutually commuting, and are
furthermore degenerate, with $P_{2\,i}{}^j=P_{3\,i}{}^j+u^aT^a_i{}^jP_1$.
As a result, the generic solution preserves at most $1/4$ of the
supersymmetries, with $1/2$ also possible in special cases (when some of
the fields are not active).  Note, however, that preservation of
supersymmetry demands the additional requirement
\begin{equation}
(\hat\gamma^m\partial_mu^aT^a_i{}^j)\epsilon_j=0,
\label{eq:addreq}
\end{equation}
which is trivially satisfied only in the rigid case.  In fact, the rigid
case is particularly simple; so long as $\epsilon_i$ is projected out by
(\ref{eq:projs}), the surviving requirement on $\epsilon_i$ for it to be a
Killing spinor is simply the parallel spinor equation
\begin{equation}
\hat\nabla_m\epsilon_i=0.
\end{equation}
In this case, the base has SU(2) holonomy, and the solution is either
$1/2$ or $1/4$ supersymmetric, depending on the set of active fields.

In the non-rigid case, however, the situation is rather more involved.
For $\epsilon_i$ to be a Killing spinor, it must not only be projected
out by (\ref{eq:projs}), but must also satisfy the sigma model requirement
(\ref{eq:addreq}).  Provided this is the case, the content of the supersymmetry
variations (\ref{eq:timekses}) reduces to
\begin{equation}
\bigl[\hat\nabla_m\delta_i{}^j-\ft12\partial_mu^aT^a_i{}^j\bigr]\varepsilon_j=0,
\label{eq:modifkse}
\end{equation}
where $\varepsilon_i=(\mathcal H_1\mathcal H_2^2)^{1/6}\epsilon_i$.
It is easily shown that integrability of this Killing spinor equation gives
rise to an Einstein equation identical to (\ref{eq:wseins}).  In order
to count the number of preserved supersymmetries, we have to identify
USp(4) symplectic-Majorana spinors $\epsilon_i$ which simultaneously
satisfy the conditions given above.  Generically, (\ref{eq:addreq})
may be considered as a sum of four terms, one for each direction on
the base ($m=1,2,3,4$).  Schematically, the Killing spinor condition is
then of the form $\pm a\pm b\pm c\pm d=0$, with all possible combinations
of signs.  With 16 possibilities, and the observation that if one choice
of signs satisfies this condition, then the completely opposite choice
would too, we see that this generically yields a $1/8$ supersymmetric
projection.  Combining (\ref{eq:addreq}) with any single projection
from (\ref{eq:projs}) leaves the solution $1/8$ supersymmetric, while
combining this with two projections gives a solution that is $1/16$
supersymmetric ({\it i.e.}\ with a single supersymmetry out of
the original 16).

Although we have not done so, it would be noteworthy to tabulate all
possible fractions of preserved supersymmetries.  This would entail
a somewhat more sophisticated investigation of (\ref{eq:addreq}) to
identify special cases away from the generic $1/8$ fraction of supersymmetry
and to ensure their compatibility with the projections of (\ref{eq:projs}).
(Kinematically, the projection (\ref{eq:addreq}) alone gives the fractions
$0,1,2,3,4,8$ out of $8$.  However, it remains to be seen
whether all such possibilities can be realized.)
In this respect, the tools of generalized holonomy
\cite{Duff:2003ec,Hull:2003mf} may also be useful in enumerating the
possibilities.

To summarize, the supersymmetric time-like solutions with SU(2)
structure are given by the bosonic fields
\begin{eqnarray}
&&ds^2=-(\mathcal H_1\mathcal H_2^2)^{-2/3}(dt+\omega)^2
+(\mathcal H_1\mathcal H_2^2)^{1/3}h_{mn}dx^mdx^n,\nonumber\\
&&G=-d[\mathcal H_1^{-1}(dt+\omega)]-G_1^+,\qquad
F^a=d[u^a\mathcal H_2^{-1}(dt+\omega)]+u^aG_2^+,
\nonumber\\
&&e^{\fft3{\sqrt6}\phi}=\mathcal H_2/\mathcal H_1,
\end{eqnarray}
where self-duality (the $+$ superscript) is with respect to the four-dimensional
base metric $h_{mn}$.  The solution is specified by the set of functions
(fields)
\begin{equation}
u^a,\qquad\mathcal H_1,\qquad\mathcal H_2,\qquad G_1^+,\qquad
G_2^+,\qquad h_{mn},
\end{equation}
which satisfy the Bianchi identities (\ref{eq:su2bianchi})
\begin{equation}
dG_1^+=0.\qquad d(u^aG_2^+)=0,
\end{equation}
scalar equations of motion (\ref{eq:scaeom})
\begin{equation}
\Box_4\mathcal H_1=\ft12(G_2^+)^2,\qquad
(\Box_4-\hat R)\mathcal H_2=\ft12G_1^+G_2^+,\qquad
(\Box_4-\hat R)u^a=0,
\end{equation}
Einstein equation on the base (\ref{eq:wseins})
\begin{equation}
\hat R_{mn}=-\partial_mu^a\partial_nu^a,
\label{eq:eins2}
\end{equation}
the relation
\begin{equation}
(d\omega)^+=-\ft12\mathcal H_1G_1^+-\mathcal H_2G_2^+,
\end{equation}
and also the sigma model supersymmetry conditions (\ref{eq:addreq}) and
(\ref{eq:modifkse}).  (Actually, the Killing spinor condition
(\ref{eq:modifkse}) implies the Einstein equation (\ref{eq:eins2}) on
the base.)

In the rigid case ($u^a=\hbox{constant}$), the base metric $h_{mn}$ has
SU(2) holonomy, and this $\mathcal N=4$ solution becomes a straightforward
generalization of the timelike $\mathcal N=2$ case analyzed in
\cite{Gauntlett:2002nw}.  Viewed from an $\mathcal N=2$ perspective, the
rigid case is essentially that of $\mathcal N=2$ supergravity coupled with
a single vector multiplet.  This results in a `two-charge' extension of
the `one-charge' (graviphoton only) solution given in \cite{Gauntlett:2002nw},
and is the origin of the second harmonic function $\mathcal H_2$ along with
a second self-dual two-form $G_2^+$.  From an $\mathcal N=2$
point of view, these solutions preserve either $0$, $1/2$ or all of the
supersymmetries, while under $\mathcal N=4$ they may preserve either $0$,
$1/4$, $1/2$ or all of the supersymmetries.

Of course, in the non-rigid case, additional fractions (such as $1/16$ and
$1/8$) are also allowed.  To better understand this non-rigid case, we
note that the $\mathcal N=4$ supergravity multiplet
\begin{equation}
(g_{\mu\nu},A_\mu^{[ij]|},B_\mu,\phi,\chi^i,\psi_\mu^i)
\end{equation}
admits the decomposition into an $\mathcal N=2$ supergravity multiplet
coupled to one vector and one gravitino multiplet
\begin{equation}
(g_{\mu\nu},A_\mu,\psi_\mu^i)+(A_\mu,\phi,\chi^i)
+(A_\mu^{iI},\chi^I,\psi_\mu^I)
\end{equation}
(where $i=1,2$ and $I=1,2$).  The graviphoton along with the vector in
the vector multiplet is a linear combination of $B_\mu$ and $u^aA_\mu^a$
({\it i.e.}\ the component of $A_\mu^a$ along $u^a$).  These two U(1)
fields carry electric components characterized by $\mathcal H_1$ and
$\mathcal H_2$ as well as magnetic components given by $G_1^+$
and $G_2^+$
\begin{equation}
-G=d[\mathcal H_1^{-1}(dt+\omega)]+G_1^+,\qquad
u^aF^a=d[\mathcal H_2^{-1}(dt+\omega)]+G_2^+.
\end{equation}
The remaining four field strengths in the gravitino multiplet are given
by projection with $\Pi_4^{ab}$
\begin{equation}
\Pi_4^{ab}F^b=du^a\wedge[\mathcal H_2^{-1}(dt+\omega)],
\end{equation}
and are only active in the non-rigid case.  Thus, from an $\mathcal N=2$
point of view, the non-rigid case corresponds to excitations of the
gravitino multiplet.  Because of this, such non-rigid solutions
are true $\mathcal N=4$ configurations without corresponding realization
within an $\mathcal N=2$ truncation.

\subsection{Time-like with SO(4) structure}

We now turn to the SO(4) structure case, which occurs when $(f^a)^2>f^2$.
As demonstrated in Section~\ref{sec:timelike}, the SO(4) valued one-forms
$V^{(4)\,a}$ define a natural vielbein basis for the metric of the form
\begin{eqnarray}
ds^2&=&-(e^0)^2+(e^a)^2\nonumber\\
&=&-(f^a)^2(dt+\omega)^2+((f^b)^2-f^2)^{-1}V_\mu^{(4)\,a}V_\nu^{(4)\,a}.
\end{eqnarray}
Recall that, although the SO(5) index $a$ runs from $1$ through $5$, the
constraint $u^aV_\mu^{(4)\,a}=0$ ensures that it only takes values in the
$\mathbf4$ of SO(4).

In order to obtain the full solution with SO(4) structure, we make
use of the fact that all spinor bilinears except $f$ and $f^a$ are fully
specified in terms of the metric and vielbein elements through (\ref{eq:viers})
and (\ref{eq:phisolv}).  In this case, we may solve directly for the
gauge fields $F^a$ and $G$ by noting that an arbitrary two-form $\mathcal F$
obeys the relation
\begin{equation}
i_K *i_K *\mathcal F=-K^2\mathcal F+K\w i_K\mathcal F.
\end{equation}
Taking $K^2=-(f^a)^2$ then allows us to write
\begin{equation}
(f^a)^2\mathcal F=-K\wedge(i_K\mathcal F)+i_K*(i_K*\mathcal F),
\end{equation}
which essentially splits $\mathcal F$ into components along $K$ and
orthogonal to $K$.

In fact, the differential identities (\ref{KG}), (\ref{KF}), (\ref{K*G})
and (\ref{K*F}) provide sufficient information for disentangling all
components of $F^a$ and $G$ through use of the above relation.  In this
manner, we obtain
\begin{eqnarray}
\kern7em&&\kern-8em e^{-\fft2{\sqrt6}\phi}(f^a)^2G
-2e^{\fft1{\sqrt6}\phi}f(f^aF^a)\nonumber\\
&=&e^{-\fft2{\sqrt6}\phi}K\wedge d(e^{\fft2{\sqrt6}\phi}f)
-e^{\fft2{\sqrt6}\phi}i_K*d(e^{-\fft2{\sqrt6}\phi}K)
+2e^{-\fft1{\sqrt6}\phi}f^ad(e^{\fft1{\sqrt6}\phi}V^a),\nonumber\\
&&\kern-8em e^{-\fft2{\sqrt6}\phi}ff^aG
+e^{\fft1{\sqrt6}\phi}[((f^a)^2-f^2)\delta^{ab}-2f^af^b]F^b\nonumber\\
&=&e^{\fft1{\sqrt6}\phi}K\wedge d(e^{-\fft1{\sqrt6}\phi}f^a)
-e^{-\fft1{\sqrt6}\phi}i_K*d(e^{\fft1{\sqrt6}\phi}V^a)
+e^{\fft2{\sqrt6}\phi}f^ad(e^{-\fft2{\sqrt6}\phi}K)\nn\\
&&+e^{-\fft1{\sqrt6}\phi}fd(e^{\fft1{\sqrt6}\phi}V^a).
\end{eqnarray}
Solving this for $F^a$ and $G$ gives
\begin{eqnarray}
((f^a)^2-f^2)(f^b)^2G&=&2f\Bigl[e^{\fft1{\sqrt6}\phi}f^ai_K
*d(e^{\fft1{\sqrt6}\phi}V^a)
-e^{\fft4{\sqrt6}\phi}(f^a)^2d(e^{-\fft2{\sqrt6}\phi}K)\nonumber\\
&&\qquad
-e^{\fft3{\sqrt6}\phi}f^aK\wedge d(e^{-\fft1{\sqrt6}\phi}f^a)\Bigr]
+2e^{\phi/\sqrt6}(f^a)^2f^bd(e^{\fft1{\sqrt6}\phi}V^b)\nonumber\\
&&+((f^a)^2+f^2)\left[K\wedge d(e^{\fft2{\sqrt6}\phi}f)
-e^{\fft4{\sqrt6}\phi}i_K*d(e^{-\fft2{\sqrt6}\phi}K)\right],\nonumber\\
((f^b)^2-f^2)(f^c)^2F^a&=&f\Bigl[
-e^{\fft1{\sqrt6}\phi}f^ai_K*d(e^{-\fft2{\sqrt6}\phi}K)
+e^{-\fft2{\sqrt6}\phi}(f^b)^2d(e^{\fft1{\sqrt6}\phi}V^a)\nonumber\\
&&\qquad+e^{-\fft3{\sqrt6}\phi}f^aK\wedge d(e^{\fft2{\sqrt6}\phi}f)\Bigr]
-e^{\fft1{\sqrt6}\phi}f^a(f^b)^2d(e^{-\fft2{\sqrt6}\phi}K)\nonumber\\
&&+(2f^af^b-\delta^{ab}(f^c)^2)\Bigl[-K\wedge d(e^{-\fft1{\sqrt6}\phi}f^b)
+e^{-\fft2{\sqrt6}\phi}i_K*d(e^{\fft1{\sqrt6}\phi}V^b)\Bigr].\nonumber\\
\end{eqnarray}
By decomposing the vector $V^a$ according to (\ref{eq:vdecomp}), we
finally arrive at the expressions
\begin{eqnarray}
((f^a)^2-f^2)(f^b)^2G&=&-((f^a)^2-f^2)\left[
K\wedge d(e^{\fft2{\sqrt6}\phi}f)+e^{\fft2{\sqrt6}\phi}i_K*dK\right]
\nonumber\\
&&+2e^{\fft2{\sqrt6}\phi}\left[(f^b)^2(V^{(4)\,a}\wedge df^a)
+fi_K*(V^{(4)\,a}\wedge df^a)\right],\nonumber\\
((f^b)^2-f^2)(f^c)^2F^a&=&((f^c)^2-f^2)\left[(f^c)^2K\wedge
d(e^{-\fft1{\sqrt6}\phi}f^a(f^c)^{-2})+e^{-\fft1{\sqrt6}\phi}f^adK\right]
\nonumber\\
&&+2e^{-\fft1{\sqrt6}\phi}f^ai_K*(V^{(4)\,b}\wedge df^b)\nonumber\\
&&+e^{-\fft2{\sqrt6}\phi}(f^b)^2\left[fd(e^{\fft1{\sqrt6}\phi}V^{(4)\,a})
-i_K*d(e^{\fft1{\sqrt6}\phi}V^{(4)\,a})\right].
\label{eq:so4gauge}
\end{eqnarray}
Note that, just as in Section~\ref{sec:timelike}, these expressions
become trivial when $(f^a)^2=f^2$.

To obtain a complete solution, we must demand that the Bianchi identities
and equations of motion hold for the gauge fields given by
(\ref{eq:so4gauge}).  We have left this as an exercise to the ambitious
reader.  Nevertheless, we expect the procedure to be similar to that of
the SU(2) structure case, and hence we expect to obtain second order
equations of a form similar to (\ref{eq:su2eom}).  Note, however, that
here a decomposition of the magnetic components of $F^a$ and $G$ into
self-dual and anti-self-dual components on the base does not appear
natural; instead the Hodge duality in (\ref{eq:so4gauge}) implies
something more along the lines of taking $\widetilde{\mathcal F}=
(f+|f^a|*_4)\mathcal F$, which is not a projection.

In addition, we must still ensure that the remaining differential
identities are satisfied.  Presumably this will lead to a sigma
model equation identical to (\ref{eq:so4sigma}) for the unit-norm
vector $u^a=f^a/|f^a|$, as well as first order conditions of the
form (\ref{eq:addreq})
\begin{equation}
(\hat\gamma^m\partial_mu^aT^a_i{}^j)\epsilon_j=0.
\end{equation}
{}From this point of view, the supersymmetry analysis of the SO(4)
structure case is rather similar to that of the SU(2) structure case
given above.  A potentially important distinction, however, is that
in the present case the Killing spinor $\epsilon_i$ does {\it not} satisfy
the simple time-direction projection $P_1\epsilon_i=0$ with the SU(2)
structure projection $P_1$ given by (\ref{eq:projs}).  (A simple
way to see this is to realize that $P_1\epsilon_i=0$ implies that
$K^{\overline\mu}$ points only in the $0$ direction, and that this in
turn gives $K^2=-f^2$.  When combined with the Fierz identity $K^2=-(f^a)^2$,
one obtains the SU(2) structure case $f^2=(f^a)^2$.)  As a result, the
counting of supersymmetries will presumably differ from that of the SU(2)
structure case.

\section{The null case}
\label{sec:nullsoln}

In this section we study the implications of having a null Killing
vector, and in particular use the differential identities to construct
the general class of supersymmetric backgrounds with $\mathbb R^3$
structure.  We first observe that, since in this case all scalar bispinors
vanish ($f=f^a=0$), the differential identity (\ref{K*G}) reduces to
\bea
d(\nt K)=i_K(\nf *G).
\eea
Contracting with $K^\mu$ in turn implies that
\bea
K \cdot dK=0. \label{null1}
\eea
Moreover, with $f=0$, we have  $i_K G=0$ from (\ref{KG}). Thus
\bea
K\wedge dK=0. \label{null2}
\eea
We now infer from (\ref{null1}) and (\ref{null2}) that the Killing vector
$K^\mu$ is such that it is hypersurface-orthogonal and may be written
as
\bea
K_\mu dx^\mu = H^{-1} du, \qquad K^\mu \partial_\mu=\partial_v,
\eea
where we have parametrized the five-dimensional spacetime in terms of
the coordinates $(u,v,y^m)$ with $m=1,2,3$. The coordinate $v$ is the
affine parameter along the geodesics of constant $u$.  In particular,
the five-dimensional metric can be written as
\bea
ds^2=H^{-1}({\cal F} du^2+2\,du\,dv)+H^2h_{mn}(dy^m+a^mdu)(dy^n+a^ndu).
\label{eq:nulmet}
\eea
Given that $\partial_v $ is an isometry generator, all the functions that
appear in the metric are $v$-independent.  For later convenience, we note
that this metric admits a natural vielbein basis
\bea
e^+ = H^{-1}du,\qquad e^-=dv+\ft12\mathcal Fdu,\qquad
e^{\bar m}=H\hat e^{\bar m}_m (dy^m+a^mdu),
\label{eq:nulviel}
\eea
where the dreibeins $\hat e^{\bar m}$ are related to the three-dimensional
base according to
\begin{equation}
\hat e^{\bar m}_m\hat e^{\bar m}_n=h_{mn}.
\end{equation}
Furthermore, although a
$u$ dependent coordinate transformation may be used to eliminate the
shift vectors $a^m$, just as in \cite{Gauntlett:2002nw} we find it useful
to keep this metric general, at least for the moment.

We now recall some of the results derived in Section~\ref{sec:null}
for $\mathbb R^3$ structure, namely that in the null case the 1-forms
$V^a$ as well as the 2-forms $\Phi^{ab}$ are all aligned with $K$
\bea
V^a=u^a K,\qquad \Phi^{ab}=K\wedge X^{ab}.
\eea
In order to construct the supersymmetric solutions of ${\cal N}=4$
supergravity characterized by a null Killing vector, we must go beyond
the structure equations and use the differential identities tabulated
in Appendix~B to express the solutions in terms of the spinor bilinears,
and then to solve for as many of the bispinors as possible.

From (\ref{KG}) and (\ref{KF}), we find the gauge field strengths of the
six abelian gauge fields are such that $i_K F^a=i_K G=0$.  This allows us
to introduce the decomposition
\begin{eqnarray}
F^a &=&F_{+\bar m}^a e^{+}\wedge e^{\bar m} + \ft12F^a_{\bar m \bar n}
e^{\bar m}\wedge e^{\bar n},\nonumber\\
G&=&G_{+\bar m}e^{+} \wedge e^{\bar m}+\ft12G_{\bar m\bar n} e^{\bar m}
\wedge e^{\bar n}.
\end{eqnarray}
Furthermore, the components $F_{\bar m\bar n}^a$ and $G_{\bar m\bar n}$
lying on the three-dimensional base can be found from
the ($m+$) components of (\ref{K*G}) and (\ref{V*F}). Concretely, we obtain
\bea
\hat F_{mn}^a&=&H^{-2} \epsilon_{mn}{}^p
(u^a \partial_p \mathcal H_2-\mathcal H_2 \partial_p u^a),\nonumber\\
\hat G_{mn}&=&-H^{-2}\epsilon_{mn}{}^p\partial_p \mathcal H_1,
\label{eq:nfg}
\eea
where the hatted quantities are defined with respect to the three-dimensional
base
\begin{equation}
\hat F^a_{mn}\equiv \hat e^{\bar m}_m\hat e^{\bar n}_n F^a_{\bar m\bar n},
\qquad\hat G_{mn}\equiv\hat e^{\bar m}_m\hat e^{\bar n}_nG_{\bar m\bar n}.
\end{equation}
The new functions ${\cal H}_1$, ${\cal H}_2$ showing up in (\ref{eq:nfg})
are defined as
\bea
{\cal H}_1=\2 H,\qquad {\cal H}_2=\no H.
\eea
Enforcing the Bianchi identities leads to the second order equations
\bea
\Box_3 {\cal H}_1=0,\qquad
u^a \Box_3 {\cal H}_2-{\cal H}_2\Box_3u^a=0,
\label{eq:nul2ord}
\eea
as well as the constraints
\bea
\fft1{\sqrt{h}}\partial_u(\sqrt{h}h^{mn}\partial_n\mathcal H_1)
&=&-\epsilon^{mnp}\partial_n(\hat G_{+p}+\epsilon_{pq}{}^ra^q
\partial_r\mathcal H_1),\nonumber\\
\fft1{\sqrt{h}}\partial_u(\sqrt{h}h^{mn}(u^a\partial_n\mathcal H_2
-\mathcal H_2\partial_nu^a))&=&\epsilon^{mnp}\partial_n(\hat F_{+p}^a
-\epsilon_{pq}{}^ra^q(u^a\partial_r\mathcal H_2-\mathcal H_2\partial_ru^a)),
\label{eq:nulbia}
\eea
where $\hat F^a_{+m}\equiv\hat e^{\bar m}_mF^a_{+\bar m}$ and $\hat G_{+m}
\equiv\hat e^{\bar m}_mG_{+\bar m}$.  The second order equations
(\ref{eq:nul2ord}) demonstrate that $\mathcal H_1$ is harmonic as a
function of $y^m$.  At the same time, the equation for $\mathcal H_2$
decomposes into the system
\begin{equation}
\Box_3\mathcal H_2=\mathcal H_2u^a\Box_3u^a,\qquad
\Box_3 u^a=u^a u^b\Box_3 u^b.
\label{o5 null}
\end{equation}
Note the similarity with the corresponding equations (\ref{eq:su2eom}) and
(\ref{eq:so4sigma}) in the timelike case.  In particular, this
reveals that the $u^a$'s define an O(5) vector model (this time
on the three-dimensional base as opposed to a four-dimensional base in
the timelike case).

The equations of motion for the field strengths provide additional
constraints on the null components $\hat F^a_{+m}$ and $\hat G_{+m}$.
However we defer these to later, and instead focus first on the one-forms
$X^{ab}$ on the three-dimensional base.  As in the timelike case, these
turn out to be closely related to the behavior of the O(5) vector $u^a$.
Starting from (\ref{VG}), we see that $\Phi^{ab}\wedge F^b=0$, which
yields the condition
\begin{equation}
X_m^{ab}h^{mn}\partial_nu^b=0.
\end{equation}
Furthermore, from (\ref{V*G}), we find the relation
\be
du^a=-*_3 X^{ab} \wedge du^b, \label{du null}
\ee
where $*_3$ is defined with respect to the metric $h_{mn}$.  In
addition, the ($+[mn]$) component of the differential identity obeyed by
the two-form $\Phi^{ab}$ (\ref{eq:dPhi}) yields
\bea
d X^{ab}=2 *_3 u^{[a} du^{b]}.
\eea
The previous two equations can be combined into
\bea
dX^{ab}+\mathcal A^{ac}\wedge X^{cb}+\mathcal A^{bc}\wedge X^{ac}=0,
\qquad \mathcal A^{ab}=2u^{[a} du^{b]},
\label{dX}
\eea
where we have introduced the composite O(5) connection $\mathcal A^{ab}$
(\ref{eq:composite}).  Notice that, in contrast to the minimal
$\mathcal N=2$ supergravity, the one-forms $X^{ab}$ are not generically
closed ({\it i.e.}\ for non-trivial $u^a$ configurations).  A bit more work
is required to extract
\bea
\hat\nabla_m X^{ab}_n=2X^{c[a}_{(m} u^{b]}_{\vphantom{(m)}}
\partial_{n)}^{\vphantom{[a]}}
u^{c\vphantom{[]}}_{\vphantom{(m)}}+\epsilon_{mn}{}^pu^{[a}\partial_pu^{b]},
\label{eq:bmw}
\eea
from the same (\ref{eq:dPhi}). Using (\ref{du null}), we obtain the direct
analog of (\ref{eq:compcovar})
\bea
D_m^{\vphantom{a}} X^{ab}_n\equiv \hat\nabla_m^{\vphantom{a}}
X^{ab}_n+\mathcal A^{ac}_m X^{cb}_n+\mathcal A^{bc}_m X^{ac}_n=0.
\label{comp con cov deriv}
\eea

We now return to the null components of the field strengths.  The conditions
of interest follow most directly from $\nabla_+ V_+^a$:
\bea
\nabla_+u^a={\cal H}_2^{-1}X_{\bar m}^{ab}F_{+\bar m}^b
\label{f+m xm}
\eea
and from $\nabla_+ \Phi^{ab}_{+\bar m}$:
\bea
\nabla_+\Phi^{ab}_{+\bar m}&=&\nabla_{+}X_{\bar m}^{ab}-
\hat e^m_{\bar m}\partial_{[m}a_{n]}X^{n\, ab}\nn\\
&=&\frac 12 \epsilon_{\bar m\bar n\bar p}
\left[{\cal H}_1^{-1} X_{\bar n}^{ab}G_{+\bar p}-2
{\cal H}_2^{-1}\left(2u^{[a} X_{\bar n}^{b]c}+u^c X_{\bar n}^{ab}\right)
F_{+\bar p}^c\right]
-2{\cal H}_2^{-1} u^{[a}F_{+\bar m}^{b]}.\qquad
\label{nabla+ phi+m}
\eea
Note that $\nabla_+=\partial_u-a^m\hat\nabla_m$.
We decompose $\hat F_{+m}^a$ into SO(4) components according to
\begin{equation}
\hat F_{+m}^a=u^a\hat F_{+m}+\hat F_{+m}^{(4)\,a},
\end{equation}
where $\hat F_{+m}=u^a\hat F_{+m}^a$ and $\hat F_{+m}^{(4)\,a}
=\Pi_4^{ab}\hat F_{+m}^b$.  In this case, (\ref{f+m xm}) and
(\ref{nabla+ phi+m}) give rise to
\begin{eqnarray}
X^{m\,ab}(\mathcal H_2^{-1}\hat F_{+m}^{(4)\,b})&=&\nabla_+u^a,\nonumber\\
\mathcal H_1^{-1}\hat G_{+m}-2\mathcal H_2^{-1}\hat F_{+m}
&=&-\epsilon_m{}^{np}\left[\partial_na_p+\ft14X_n^{ab}\partial_uX_p^{ab}
\right].
\label{g+m f+m}
\end{eqnarray}
The SO(4) singlet term was given by multiplying both sides of
(\ref{nabla+ phi+m}) with $X^{ab}_{\bar q}$ and using the relation
\bea
X_{m}^{ab} X_{n}^{ab}=4h_{mn},
\eea
which follows from (\ref{eq:x multiplication}).  In addition, we have
used the fact that $X_p^{ab}\hat\nabla_mX_n^{ab}=0$ (which follows from
contracting (\ref{eq:bmw}) with $X_p^{ab}$) to write
$X_n^{ab}\nabla_+X_p^{ab}=X_n^{ab}\partial_uX_p^{ab}$ in (\ref{g+m f+m}).

In contrast to the null solution of $\mathcal N=2$ supergravity
\cite{Gauntlett:2002nw}, here the null components of the field strengths
are only incompletely determined.  Additional requirements on these
components may be obtained from the two-form equations of motion.  With some
manipulation, the equations of motion (\ref{eq:2eom}) give rise to
\begin{eqnarray}
\hat\nabla^m(\mathcal H_1^{-1}\hat G_{+m})&=&
-2(\mathcal H_{-1}\hat G_{+m}+\mathcal H_2^{-1}\hat F_{+m})h^{mn}
\partial_n\log\mathcal H_2+2\mathcal H_2^{-1}\hat F_{+m}^ah^{mn}\partial_nu^a,
\nonumber\\
\hat\nabla^m(\mathcal H_2^{-1}\hat F_{+m}^a)&=&
-(u^a\mathcal H_1^{-1}\hat G_{+m}+\mathcal H_2^{-1}\hat F_{+m}^a)h^{mn}
\partial_n\log\mathcal H_2+\mathcal H_1^{-1}\hat G_{+m}h^{mn}\partial_nu^a.
\label{eq:nuleom}
\end{eqnarray}
Note that subtracting twice the SO(5) singlet component of the $\hat F_{+m}^a$
equation from the $\hat G_{+m}$ equation gives rise to the divergence free
condition
\begin{equation}
\hat\nabla^m(\mathcal H_1^{-1}\hat G_{+m}-2\mathcal H_2^{-1}\hat F_{+m})=0.
\end{equation}
Combining this with (\ref{g+m f+m}) yields the consistency requirement
\begin{equation}
\epsilon^{mnp}\hat\nabla_m(X_n^{ab}\partial_uX_p^{ab})=0.
\end{equation}
It turns out, however, that this is automatically satisfied based on the
properties of $u^a$ and $X_m^{ab}$.  This allows us to conclude that
the right hand side of the second expression in (\ref{g+m f+m}) may be
written as a pure curl, as it is automatically divergence free.  Because
the shift vectors $a^m$ were introduced purely as a convenience, we may
thus absorb the somewhat awkward term $\fft14X_m^{ab}\partial_uX_p^{ab}$
into a redefinition of $a^m$.  This then gives us
\begin{equation}
\mathcal H_1^{-1}\hat G_{+m}-2\mathcal H_2^{-1}\hat F_{+m}=-\epsilon_m{}^{np}
\partial_na_p,
\label{eq:newa}
\end{equation}
which is analogous to the corresponding expression in the null $\mathcal N=2$
case \cite{Gauntlett:2002nw}.
Finally, for completeness, we note that the projection of (\ref{nabla+ phi+m})
onto the anti-self-dual $\rm SU(2)_-$ in SO(4) gives the condition
\begin{equation}
\left[\Pi_4^{ac}\Pi_4^{bd}-\ft14X_n^{ab}X^{n\,cd}\right]\nabla_+X_m^{cd}=0.
\end{equation}

To summarize what we have obtained for the field strengths,
the $\hat G_{+m}$ and $\hat F_{+m}$ components cannot be solved
for independently.  Instead, a linear combination of the two is determined
via (\ref{eq:newa}).  This is similar to what happens for the magnetic
components in the timelike case, where (\ref{eq:ofg}) demonstrates that
$\overline{G}^+$ and $\overline{F}^+$ only enter through the combination
$\mathcal H_1\overline{G}^+-2\mathcal H_2\overline F^+$.  The components
of $\hat F^{(4)\,a}_{+m}$ taking values in the $\mathbf 4$ of SO(4) are
determined only so far as their projection onto $X_m^{ab}$, as given in
(\ref{g+m f+m}).  Of course, in all cases, the Bianchi identities
(\ref{eq:nulbia}) and equations of motion (\ref{eq:nuleom}) still need to
be satisfied.

Turning now to the Killing spinor equations, from the dilatino
supersymmetry variation we find the projectors
\be
\gamma^{+}\epsilon=0,\qquad (1-u^a T^a)\epsilon=0,
\ee
as well as the constraint
\bea
\hat\gamma^m \partial_m u^a T^a \epsilon=0,
\eea
which coincides with (\ref{eq:addreq}) in the timelike case.
The supersymmetry variation of the gravitino yields one more constraint
\bea
\hat\nabla_m \epsilon=\ft12\partial_m u^a T^a \epsilon,
\eea
which also has a direct analog in the timelike case, namely
(\ref{eq:modifkse}).
The integrability conditions which follow from this equation
are the same as those derived from the covariant derivative $D_m$
defined in (\ref{comp con cov deriv}). Namely, we find that
\bea
R_{mn}=-\partial_m u^a \partial_n u^a.
\label{einstein-o5 null}
\eea
The equation (\ref{einstein-o5 null}), together with (\ref{o5 null}), can be interpreted as the Einstein equation of a three-dimensional O(5) vector model
coupled to gravity.  (However, just as in the timelike case, this model has
an unconventional sign for the stress tensor.)

If the $u^a$'s are taken to be rigid O(5) vectors, then the three-dimensional
base is not only Ricci-flat, as indicated by (\ref{einstein-o5 null}),
but is actually flat. This can be derived from (\ref{dX}); with the
one-forms $X^{ab}$ closed, we can choose coordinates
on the base such that $dy^m$ are identified with the three independent
one-forms $X^{ab}$. That these independent one-forms define a dreibein
basis follows from the multiplication rule (\ref{eq:x multiplication})
obeyed by $X^{ab}$.  The situation is rather more involved for
the non-rigid case.  For one thing, most quantities can then be functions
of the null coordinate $u$.  In this case, a slight simplification may
arise by setting the vectors $a^i=0$ through an appropriate choice of
coordinates.  Nevertheless, a complete analysis of the non-rigid case
appears somewhat formidable, and still remains to be completed.

Finally,
for solutions in the null category, it should be noted that the $R_{++}$
Einstein equation remains to be solved independently of the supersymmetry
conditions.  For the $\mathcal N=4$ model, this component of the
Einstein equation turns out to be
\bea
R_{++}+\f{1}{4}e^{\fft2{\sqrt6}\phi}F_{+\rho}^{ij}
{F_{+}}^{\rho\;ij}+\f{1}{2}e^{-\fft4{\sqrt6}\phi}G_{+\rho}{G_+}^{\rho}
+\f{1}{2}\p _+\phi\p _+\phi=0.
\label{eq:eins}
\eea
Given the null metric (\ref{eq:nulmet}) with vielbeins (\ref{eq:nulviel}),
we find the expression for $R_{++}$
\be
R_{++}=-\f{1}{2H}\Box_3{\cal F}-H\nabla_+W_{\bar m\bar m}-
W_{(\bar{m}\bar{n})}W_{(\bar{m}\bar{n})},
\ee
where
\be
W_{\bar{m}\bar{n}}=\nabla_+H\d_{\bar{m}\bar{n}}+H\d
_{\bar{m}\bar{p}}(\nabla_+\hat{e}^{\bar{p}}_m)\hat{e}^{m}_{\bar{n}}-H\d
_{\bar{m}\bar{p}}\hat{e}^{\bar{p}}_m\hat{e}^n_{\bar{n}}\hat\nabla_na^m.
\ee
The actual Einstein equation, (\ref{eq:eins}), is rather cumbersome as
the null components $\hat F_{+m}^a$ and $\hat G_{+m}$ are only partially
determined in the present analysis.

\section{Solutions}

As discussed above in Section~\ref{sec:timesolnsu2},
the field content of ${\cal N}=4$ five dimensional supergravity can be 
decomposed in ${\cal N}=2$ representations as follows: the minimal
supergravity multiplet [the metric, one gauge field, and two gravitini
transforming in the $\mathbf2$ of USp(2)], one vector multiplet (one gauge
field and  one scalar, the dilaton) and a gravitino multiplet (the remaining
two gravitini and four gauge fields). 
Thus, by setting the matter multiplets to zero, we shall reproduce the 
supersymmetric solutions
of minimal five-dimensional supergravity found in \cite{Gauntlett:2002nw}.
To do so requires rigid SO(5)/SO(4) vectors $u^a$. 
Furthermore, truncating the set of gauge fields must be done such that
$i)$ for the SU(2) structure case we demand $G_1^+=G_2^+$; or
$ii)$ for the null case $F^{(4)\,a}$ must vanish.
Lastly, setting the dilaton to zero, which amounts to
${\cal H}_1={\cal H}_2$, leads to the set of equations and constraints
which determine the supersymmetric backgrounds of minimal five-dimensional
supergravity with either SU(2) holonomy or $\mathbb R^3$ structure,
respectively \cite{Gauntlett:2002nw}.

If, on the other hand, we impose the conditions that $u^a$ is rigid
but allow $\mathcal H_1$ and $\mathcal H_2$ (as well as $G_1^+$ and
$G_2^+$ in the timelike case) to be independent, then
we fall back onto the two-charge solutions of minimal supergravity coupled 
to one vector multiplet described in \cite{Gutowski:2004yv}.
In this class of rigid solutions, we are also able to reproduce a subset
of the black ring solutions of \cite{Bena:2004de}, which are
characterized by two electric and two (magnetic) dipole charges.
To see this, select the case of a time-like Killing vector and begin again
by choosing rigid $u^a$. Then simply identify the
three harmonic functions of \cite{Bena:2004de} as
$Z_1={\cal H}_1$, $Z_2=Z_3={\cal H}_2$; these harmonic functions
determine the electric charge distributions. The magnetic fields
of \cite{Bena:2004de} are to be identified with
$G_1=G_1^+$, $G_2=G_3=G_2^+$.

Notice that in all the previous examples we began by selecting a rigid
five-dimensional unit norm vector $u^a$. As discussed in Section 3.1,  
having a non-trivial $u^a$ amounts
to turning on the gravitino multiplet. In this case, the starting point
in constructing the five-dimensional supersymmetric backgrounds must be 
solving a gravitating SO(5) vector sigma model, in three or four dimensions. 
The worldvolume of the sigma model is a Riemannian manifold (positive 
definite metric).  
We proceed next to construct a few solutions of the gravitating vector model
\be
\Box u^a=R u^a, \qquad R_{mn}=-\partial_m u^a\partial_n u^a.\label{grav vector}
\ee
At the same time, to ensure that these solutions lead to five-dimensional
backgrounds, we must enforce the supersymmetry constraint
\be
\gamma^m \partial_m u^a T^a\epsilon=0.
\ee
The simplest case has the  $u^a$'s defining maps from a one-dimensional
manifold into a circle. However, this is at odds with the supersymmetry
constraint, as $\gamma^1 T^1$ has no zero eigenvalues.
The first non-trivial case corresponds to maps from a two-dimensional 
manifold 
\be
ds^2=e^{2\Psi(z,\bar z)}dzd\bar z
\ee
into a two-sphere $S^2$. These maps define the stereographic projection
\be
u^a=\bigg(\frac{\psi+\bar\psi}{1+\psi\bar\psi},i\frac{\psi-\bar\psi}{
1+\psi\bar\psi}, \frac{1-\psi\bar\psi}{1+\psi\bar\psi},0,0\bigg),
\ee
where we have assumed that $\psi=\psi(z)$ are holomorphic functions.
The supersymmetry constraint is satisfied since
\be
\partial u^a T^a=\frac{\partial \psi}{(1+\psi\bar\psi)^2}
\left((1-\bar \psi^2)T^1+i(1+\bar\psi^2)T^2+2\bar\psi T^3\right),
\ee
and the $SO(5)$ matrix which appears between the brackets has zero 
eigenvalues. The $\gamma^{\bar z}\bar u^a T^a \epsilon$ term vanishes
because $\gamma^{\bar z}$ has zero eigenvalues.
The solution to (\ref{grav vector}) yields 
\be
e^{2\Psi}=(1+\psi\bar\psi)^2|\xi(z)|^2,
\ee
where $\xi(z)$ is an arbitrary holomorphic function of $z$.

To construct the corresponding five-dimensional solution, we first 
extend the two-dimensional base to a three or four-dimensional manifold.
Then we need to solve for the 
functions ${\cal H}_1, {\cal H}_2$ such that
$\Box{\cal H}_1=0$ and $\Box {\cal H}_2=R{\cal H}_2$ in the null case,
as well as in the SU(2) structure case in the absence of fluxes.
Recall that in the timelike case the warp factor of the 
five-dimensional metric is
$f=\left({\cal H}_1{\cal H}_2{}^2\right)^{-1/3}$, 
with $ds_5^2=-f^2 dt^2 +f^{-1} ds_4^2$, while in the null case the warp
factor is 
$H=\left({\cal H}_1{\cal H}_2{}^2\right)^{1/3}$, with 
$ds_5^2=H^{-1}({\cal F} du^2+2\,du\,dv)+H^2 ds_3^2$.
Since ${\cal H}_1$ is harmonic, we can take ${\cal H}_1=1$. On the other
hand, ${\cal H}_2$ satisfies the same equation as $u^a$. By identifying
${\cal H}_2$ with $u^3$ we generate a warp factor which has only a radial
dependence on the two-dimensional base. Even though the base is regular, 
the five-dimensional solution may be singular. The reason
why this could happen is that zeros of ${\cal H}_2$ translate
into singularities. In this case, the singularities of the five-dimensional
background are localized on the locus $\psi\bar\psi=1$. Noticing that the 
volume of the base manifold vanishes when  $\psi\bar\psi=1$, we conclude
that the singularity is point-like. 
Other choices of ${\cal H}_2$ (such as turning on some Fourier modes, which 
can be done by identifying ${\cal H}_2$ with $u^1$ or $u^2$) could give a 
different picture in terms of the location of the singularity, but
they cannot remove it. For instance ${\cal H}_2=u^1$ vanishes when 
$Re(\psi)=0$.

We find a similar story unfolding when considering higher-dimensional maps
from conformally flat spaces to spheres. For $u^a$ spanning an $S^3$,
\be
u^a=\Bigl(\sin(\psi(r))\sin\theta\cos\phi,\sin(\psi(r))\sin\theta\sin\phi,
\sin(\psi(r))\cos\theta,\cos(\psi(r)),0\Bigr),
\ee
and with the three-dimensional base given by
\be
ds_3^2=e^{2\Psi(r)}(dr^2+r^2d\Omega_2^2),
\ee
we find that the supersymmetry constraint yields 
\be
\left(
\gamma^r \frac{d\psi }{dr} T^1(r,\theta,\phi)+
\gamma^\theta \sin \psi T^2(r,\theta,\phi)
+\gamma^\phi
\sin \psi\sin\theta T^3(r,\theta,\phi)\right)\epsilon=0,
\ee
where $T^{1,2,3}(r,\theta,\phi)$ are $SO(4)$-rotated $SO(5)$ matrices.
Given that $[\gamma^r T^1,\gamma^\theta T^2]=0$, {\it etc.}, these matrices
can be diagonalized simultaneously. The existence of zero eigenvalues
requires either
\be
i)\quad\psi={\rm const}, \qquad\hbox{or}\qquad
ii)\quad\frac{d\psi}{dr}=\pm 2\frac{sin \psi}{r}.
\ee
Next we proceed to solve the gravitating vector sigma model
equations. We find that the second case is the only 
possibility, leading to
\be
\cos\psi=\frac{1-r^4}{1+r^4}.
\ee 
The first option, $\psi=$~constant, is excluded since spheres,
while compatible with supersymmetry, have positive curvature. 
On the other hand, we are looking for solutions to a gravitating $SO(5)$ 
sigma model with a negative contribution 
to the stress-energy tensor. Therefore we are looking for manifolds of 
negative curvature.

For maps from conformally flat four dimensional manifolds into $S^4$,
supersymmetry requires that either
\be
i)\quad\frac{d\psi}{dr}=\pm\frac{\sin\psi}{r},\qquad\hbox{or}\qquad
ii)\quad\frac{d\psi}{dr}=\pm3\frac{\sin\psi}r,
\ee
with the latter being realized as a solution of the gravitating sigma model
\be
\cos\psi=\frac{1-r^6}{1+r^6}.
\ee

As discussed before, to construct the corresponding 
five dimensional solutions requires solving for the harmonic function 
${\cal H}_1$ as well as for ${\cal H}_2$. Notice that since we may identify
${\cal H}_2$ with any of $u^a$, and since $u^a$'s are unit vectors
spanning a sphere, they will vanish: $u^4$ has zeros at $r=1$, and the rest
vanish when $r=0$. In addition $u^1$, $u^2$ and $u^3$ have zeros coming from
the angular dependence. At the location of the zeros of ${\cal H}_2$, the
five-dimensional solution will be singular.

It is worth asking whether by turning on fluxes we can improve the current 
predicament. In the null case, this will have no repercussions, since 
the equation for ${\cal H}_2$ is insensitive to any flux. In the time-like
case with SU(2) structure, it appears at first, that by adding fluxes
$G_1^+, G_2^+$ one could make a difference. However, the flux
$G_2^+$ is constrained by 
$d(u^a G_2^+)=0$. With the $u^a$'s spanning at least a two-sphere, all
components of the self-dual two-form $G_2^+$ are set to zero, and no
additional source term is generated for ${\cal H}_2$.

We have explored a few other solutions. Another simple way to generate negative
curvature spaces is to consider cones over spheres. For instance 
\be
ds^2_3= r^B (dr^2+ A r^2 d\Omega_2),\label{cone}
\ee
with $u^a$ spanning $S^2$ is compatible both with supersymmetry and 
with the gravitating vector sigma-model equations.
\be
R_{rr}=0,\qquad R_{\theta\theta}=-1, 
\ee
provided that
\be
A(B+2)^2=8,\qquad A>0.
\ee
In this case, solving for ${\cal H}_2={\cal H}_2(r)$
yields 
\be
{\cal H}_2= \mathrm{Re}\left(r^{\frac{(B+2)}4 (1\pm i\sqrt 3)}\right),
\ee
which has an infinite number of nodes. As has been explained before,
the five-dimensional solution built on the three-dimensional 
manifold (\ref{cone}) will be singular at the location of these nodes.

Lastly, we have investigated a warped three dimensional manifold
\be
ds^2_3=dy^2+y^2 e^{2\Psi(r)}(dr^2+ r^2d\phi^2)
\ee
and with $u^a=(\sin(\psi(r))\sin\phi,\sin(\psi(r))\cos\phi,\cos(\psi(r)),0,0)$.
Given that the supersymmetry constraint is satisfied, we move onto 
the gravitating vector sigma-model equations.
The warp factor $y^2$ is the only choice up to $y$-translations which 
solves
\be
R_{yy}=0
\ee
other than a trivial warping $y^0=1$.  This time ${\cal H}_2$ can be a
function of both $r$ and $y$.  We found solutions using separation of
variables, ${\cal H}_2=h(r)\tilde h(y)$.  It turns out, however, that
if $\tilde h(y)$ has no zeros, then $h(r)$ will, and {\it vice versa}.

As a final comment, we would like to mention that we have inquired about
the existence of generic SO(4) structure solutions.  Under the simple
assumptions of a rigid $f^a$ and of a flat four-dimensional base with all
fields depending on a single variable, the only solutions to the Bianchi
identities and equations of motion compatible with the supersymmetry
constraints turned out to be trivial, with $f^a=$~constant and $f=$~constant.
It remains to be seen whether there are any large classes of solutions
with SO(4) structure yet to be found.

\section*{Acknowledgments}

This work was supported in part by the US~Department of Energy under
grant DE-FG02-95ER40899.

\appendix
\section{Fierz Identities}

The determination of the structure groups, as well as the explicit
construction of the solutions, requires consideration of the algebraic
identities satisfied by the spinor bilinears.  These identities are
essentially Fierz identities, and are obtained by using the five-dimensional
Fierz relation
\be
4(\bep_1\ep_2)(\bep_3\ep_4)=(\bep_1\ep_4)(\bep_3\ep_2)
+(\bep_1\G_\rho\ep_4)(\bep_3\G^\rho\ep_2)
-\ft12(\bep_1\G_{\rho\sigma}\ep_4)(\bep_3\G^{\rho\sigma}\ep_2),
\ee
where the USp(4) indices have been hidden.

Although a great number of identities may be obtained, we only highlight
some of the more useful ones here.  Furthermore, as was done in the body
of the paper, we use a SO(5) notation for the bispinors
\begin{equation}
f,\qquad f^a,\qquad K_\mu,\qquad V_\mu^a,\qquad\Phi_{\mu\nu}^{ab},
\end{equation}
which were defined in (\ref{eq:so5bispin}).  Note that (when considered as
tangent space indices) the spacetime indices $\mu,\nu,\ldots$ take values
in SO(1,4), while indices $a,b,\ldots$ are valued in SO(5).  Because of
this similarity in groups, the Fierz identities exhibit a formal symmetry
under the interchange of spacetime and internal space indices along with
the exchange $f^a\leftrightarrow K_\mu$.

We organize the algebraic identities according to the number of open
spacetime and internal space indices.  For the scalar-singlet combination,
we have
\begin{eqnarray}
(K_\mu)^2&=&-(f^a)^2,\label{eq:KKf}\\
(V_\mu^a)^2&=&-5f^2+4(f^a)^2,\label{eq:vasq}\\
\ft18(\Phi_{\mu\nu}^{ab})^2&=&5f^2+(f^a)^2.\label{eq:phimnabsq}
\end{eqnarray}
The first identity above demonstrates that the Killing vector $K^\mu$ is
nowhere spacelike.  For the vector-singlet case, we have
\begin{eqnarray}
fK_\mu&=&f^aV_\mu^a,\label{eq:fKf}\\
fK_\mu&=&\ft1{96}\epsilon_\mu{}^{\nu\rho\lambda\sigma}\Phi_{\nu\rho}^{ab}
\Phi_{\lambda\sigma}^{ab}
\end{eqnarray}
while the scalar-$\mathbf5$ case yields
\begin{eqnarray}
ff^a&=&-K^{\mu}V_{\mu}^a,\label{eq:KVf}\\
ff^a&=&\ft{1}{96}\epsilon^{abcde}\Phi_{\mu\nu}^{bc}\Phi^{\mu\nu\,de}.
\end{eqnarray}

Turning to cases with additional open indices, we start with the
vector-$\mathbf5$ relations
\begin{eqnarray}
\ft1{96}\epsilon_\mu{}^{\nu\rho\lambda\sigma}\epsilon^{abcde}
\Phi_{\nu\rho}^{bc}\Phi_{\lambda\sigma}^{de}&=&fV_\mu^a+f^aK_\mu,\\
\ft14\Phi_{\mu\nu}^{ab}V^{\nu\,b}&=&fV_\mu^a-f^aK_\mu.
\end{eqnarray}
Next, we find that the
scalar-symmetric tensor ($\mathbf1+\mathbf{14}$) combination gives
\begin{eqnarray}
V_\mu^aV^{\mu\,b}&=&-f^af^b+\d^{ab}((f^c)^2-f^2)\label{eq:VVf}\\
\ft14\Phi_{\mu\nu}^{ac}\Phi^{\mu\nu\,bc}&=&-3f^af^b+\delta^{ab}((f^c)^2+2f^2).
\label{eq:02phiphi}
\end{eqnarray}
Note that contraction with the SO(5) invariant tensor $\delta^{ab}$ gives
the singlet identities (\ref{eq:vasq}) and (\ref{eq:phimnabsq}) above.
The flipped version of (\ref{eq:VVf}) is the tensor-singlet combination
\begin{equation}
V_\mu^aV_\nu^a=K_\mu K_\nu+g_{\mu\nu}((f^a)^2-f^2).
\label{eq:vmuavnua}
\end{equation}
Turning next to the vector-antisymmetric tensor ($\mathbf{10}$) combination,
we find
\begin{eqnarray}
K^\mu\Phi _{\mu\nu}^{ab}&=&-\ft{1}{6}\epsilon^{abcde}V^{\mu\,c}
\Phi_{\mu\nu}^{de},\\
K^\mu\Phi_{\mu\nu}^{ab}&=&-2f^{[a}V_\nu^{b]}.\label{eq:kmuphimu}
\end{eqnarray}
The latter equation has a flipped tensor-$\mathbf5$ version
\begin{equation}
f^a\Phi_{\mu\nu}^{ab}=2K_{[\mu}V_{\nu]}^b.
\label{eq:faphiab}
\end{equation}
Finally, a couple of useful tensor-$\mathbf{10}$ relations are
\begin{eqnarray}
\epsilon_{\mu\nu}{}^{\rho\lambda\sigma}K_\rho\Phi_{\lambda\sigma}^{ab}
&=&-\epsilon^{abcde}f^c\Phi_{\mu\nu}^{de},\label{eq:epskphi}\\
\epsilon_{\mu\nu}{}^{\rho\lambda\sigma}K_\rho\Phi_{\lambda\sigma}^{ab}
&=&4V_\mu^{[a}V_\nu^{b]}-2f\Phi_{\mu\nu}^{ab}.\label{eq:VwedgeVf}
\end{eqnarray}

For more complicated combinations, we do not perform a complete decomposition
into irreducible representations, but merely list the number of spacetime
and internal space indices according to (\# spacetime, \# internal).
In the $(3,1)$ and $(3,2)$ categories, we have
\begin{eqnarray}
V_{\mu}^a\P^{ab}_{\nu\lambda}&=&-2g_{\mu [\nu}(fV_{\lambda}^b-f^bK_{\lambda]})
+\epsilon_{\mu\nu\lambda}{}^{\rho\s}V_{\rho}^bK_{\s},\label{eq:vmuaphi}\\
K_{\mu}\P^{ab}_{\nu\lambda}&=&-2g_{\mu [\nu}(f^aV_{\lambda}^b-f^bV_{\lambda]}^a)
-\epsilon_{\mu\nu\lambda}{}^{\rho\s}V_{\rho}^aV^b_{\s}
+\ft{1}{2}\epsilon^{abcde}V_{\mu}^c\P^{de}_{\nu\lambda}.\label{eq:kphi}
\end{eqnarray}
These identities are useful for deducing the basic properties of the two-form
$\Phi^{ab}$.  Additional information on $\Phi^{ab}$ and its relation to
SU(2) or $\mathbb R^3$ structure can be obtained from the $(1,4)$ identity
\begin{eqnarray}
\ft{1}{4}\epsilon_\mu{}^{\nu\rho\lambda\s}\P^{ab}_{\nu\rho}
\P^{cd}_{\lambda\s}&=&
\epsilon^{abcde}(f^eK_{\mu}+fV_{\mu}^e)\nonumber\\
&&+2fK_\mu(\d^{ac}\d^{bd}-\d^{ad}\d^{bc})
-2\bigl[f^{(a}V_{\mu}^{c)}\d^{bd}+f^{(b}V_{\mu}^{d)}\d^{ac}
-f^{(a}V_{\mu}^{d)}\d^{bc}-f^{(b}V_{\mu}^{c)}\d^{ad}\bigr],\nonumber\\
\end{eqnarray}
as well as the $(2,2)$ identity
\bea
\P^{ab}_{\mu\lambda}\P^{\lambda\,bc}_{\;\;\nu}&=&\d^{ac}\bigl[3K_{\mu}K_{\nu}
+g_{\mu\nu}(f^2+2(f^d)^2)\bigr]
-3g_{\mu\nu}f^af^c-3V_{\mu}^{(a}V^{c)}_{\nu}\nn\\
&&+V_{\mu}^{[a}V_{\nu}^{c]}-2f\P^{ac}_{\mu\nu}.
\eea
A contraction on the internal indices results in the $(2,0)$ counterpart
of (\ref{eq:02phiphi})
\begin{equation}
\ft14\Phi_{\mu\lambda}^{ab}\Phi^{\lambda\,ab}_{\;\;\nu}
=-3K_\mu K_\nu-g_{\mu\nu}((f^c)^2+2f^2).
\end{equation}
The identities with more open indices are rather tedious, but useful
for completing the determination of the structure.  In the $(4,2)$
category, we have
\begin{eqnarray}
\Phi_{\mu\nu}^{ac}\Phi_{\rho\sigma}^{bc}&=&
\epsilon_{\mu\nu\rho\sigma}{}^\lambda(f^{(a}V_\lambda^{b)}
-\delta^{ab}fK_\lambda)
+(g_{\mu\rho}g_{\nu\sigma}-g_{\mu\sigma}g_{\nu\rho})(-f^af^b+\delta^{ab}f^2)
\nonumber\\
&&+\delta^{ab}\bigl[K_\mu K_\rho g_{\nu\sigma}+K_\nu K_\sigma g_{\mu\rho}
-K_\nu K_\rho g_{\mu\sigma}-K_\mu K_\sigma g_{\nu\rho}\bigr]\nonumber\\
&&+\ft12\bigl[K_{[\mu}\epsilon_{\nu]\rho\sigma}{}^{\alpha\beta}
-K_{[\rho}\epsilon_{\sigma]\mu\nu}{}^{\alpha\beta}\bigr]\Phi_{\alpha\beta}^{ab}
-f\bigl[\Phi_{\mu\rho}^{ab}g_{\nu\sigma}+\Phi_{\nu\sigma}^{ab}g_{\mu\rho}
-\Phi_{\nu\rho}^{ab}g_{\mu\sigma}-\Phi_{\mu\sigma}^{ab}g_{\nu\rho}\bigr]
\nonumber\\
&&-\bigl[V_\mu^bV_\rho^ag_{\nu\sigma}+V_\nu^bV_\sigma^ag_{\mu\rho}
-V_\nu^bV_\rho^ag_{\mu\sigma}-V_\mu^bV_\sigma^ag_{\nu\rho}\bigr],
\label{eq:42fierz}
\end{eqnarray}
while the opposite $(2,4)$ case gives a similar expression
\begin{eqnarray}
\Phi_{\mu\lambda}^{ab}\Phi^{\lambda\,cd}_{\;\;\nu}&=&
-\epsilon^{abcde}(K_{(\mu}V_{\nu)}^e+g_{\mu\nu}ff^e)
+(\delta^{ac}\delta^{bd}-\delta^{ad}\delta^{bc})(-K_\mu K_\nu+g_{\mu\nu}K^2)
\nonumber\\
&&+g_{\mu\nu}\bigl[f^af^c\delta^{bd}+f^bf^d\delta^{ac}-f^cf^d\delta{bc}
-f^bf^c\delta^{ad}\bigr]\nonumber\\
&&+\ft12(f^{[a}\epsilon^{b]cdef}-f^{[c}\epsilon^{d]abef})\Phi_{\mu\nu}^{ef}
+f\bigl[\Phi_{\mu\nu}^{ac}\delta^{bd}+\Phi_{\mu\nu}^{bd}\delta^{ac}
-\Phi_{\mu\nu}^{ad}\delta^{bc}-\Phi_{\mu\nu}^{bc}\delta^{ad}\bigr]\nonumber\\
&&+\bigl[V_\mu^cV_\nu^a\delta^{bd}+V_\mu^dV_\nu^b\delta^{ac}
-V_\mu^cV_\nu^b\delta^{ad}-V_\mu^dV_\nu^a\delta^{bc}\bigr].
\label{eq:hks}
\end{eqnarray}
%

\section{Differential identities}

Here we provide the complete set of differential identities obtained
from the action of the dilation and gravitino variations on the USp(4)
bispinors.  We recall that the USp(4) valued scalar, vector and tensor
bispinors were defined in (\ref{eq:usp4bispin}) as
\bea
f^{[ij]}&=&i\bep ^i\ep^j,\nn\\
V_{\mu}^{[ij]}&=&\bep ^i\g _{\mu} \ep ^j,\nn\\
\Phi ^{(ij)}&=&i\bep ^i\g _{\mu\nu}\ep ^j.
\eea
These may be split into irreducible SO(5) representations according to
\bea
f^{ij}&=&f\Omega^{ij}+f^a T^{a\,ij},\nn\\
V_{\mu}^{ij}&=&K_{\mu}\Omega^{ij}+V_{\mu}^aT^{a\,ij}\nn\\
\Phi_{\mu\nu}^{ij}&=&\frac{1}{2}\Phi ^{ab}_{\mu\nu}T^{ab\,ij}.
\label{eq:so5bilin}
\eea

Although the dilatino variation (\ref{eq:deltachi}) does not lead to
derivatives on $\epsilon^i$, we nevertheless consider the resulting
expressions as `differential' identities to distinguish them from
kinematical or Fierzing relations.  By taking
$\overline\epsilon^i\{1,\Gamma_\mu,\Gamma_{\mu\nu}\}\delta\chi^j=0$,
we obtain
\bea 0&=&K^{\mu}\p _{\mu}\phi,\\
0&=&V^{\mu\,a}\p _{\mu}\phi+\ft{1}{\sqrt{6}}e^{\fft1{\sqrt6}\phi}
\Phi _{\mu\nu}^{ab}F^{\mu\nu\,b},\\
0&=&e^{-\fft2{\sqrt6}\phi}\Phi _{\mu\nu}^{ab}G^{\mu\nu}
+\ft12e^{\fft1{\sqrt6}\phi}\epsilon^{abcde}\Phi_{\mu\nu}^{cd}F^{\mu\nu\,e},\\
0&=&f\p_{\mu}\phi+\ft{2}{\sqrt{6}}e^{-\fft2{\sqrt6}\phi}G_{\mu\nu}K^{\nu}
+\ft{2}{\sqrt{6}}e^{\fft1{\sqrt6}\phi}F_{\mu\nu}^{a}V^{\nu\,a},\\
0&=&f^a\p_{\mu}\phi+\ft{2}{\sqrt{6}}e^{-\fft2{\sqrt6}\phi}G_{\mu\nu}V^{\nu\,a}
+\ft{2}{\sqrt{6}}e^{\fft1{\sqrt6}\phi}F_{\mu\nu}^aK^{\nu}
-\ft{1}{2\sqrt{6}}e^{\fft1{\sqrt6}\phi}\epsilon^{\mu\nu\lambda\rho\s}
F_{\nu\lambda}^b
\Phi_{\rho\s}^{ab},\\
0&=&\Phi_{\mu\nu}^{ab}\p^{\nu}\phi+\ft{1}{2\sqrt{6}}e^{-\fft2{\sqrt6}\phi}
\epsilon_{\mu\nu\lambda\rho\s}G^{\nu\lambda}\Phi_{\rho\s}^{ab}+\ft{1}{4\sqrt{6}}
e^{\fft1{\sqrt6}\phi}\epsilon_{\mu\nu\lambda\rho\s}\epsilon^{abcde}
\Phi_{\rho\s}^{cd}F_{\nu\lambda}^e
+\ft{4}{\sqrt{6}}e^{\fft1{\sqrt6}\phi}F_{\mu\nu}^{[a}V^{\nu\,b]},\nonumber\\
\\
0&=&\ft{2}{\sqrt{6}}e^{-\fft2{\sqrt6}\phi}fG_{\mu\nu}+2K_{[\mu}\p_{\nu]}\phi
-\ft1{\sqrt{6}}e^{-\fft2{\sqrt6}\phi}\epsilon_{\mu\nu\lambda\rho\s}
G^{\lambda\rho}K^{\s}
+\ft2{\sqrt{6}}e^{\fft1{\sqrt6}\phi}f^aF_{\mu\nu}^a\nonumber\\
&&\qquad-\ft{1}{\sqrt{6}}e^{\fft1{\sqrt6}\phi}\epsilon_{\mu\nu\lambda\rho\s}
F_{\lambda\rho}^aV_{\s}^a,\\
0&=&\ft{2}{\sqrt{6}}e^{-\fft2{\sqrt6}\phi}f^aG_{\mu\nu}
+2V_{[\mu}^a\p_{\nu]}\phi-\ft{1}{\sqrt{6}}e^{-\fft2{\sqrt6}\phi}
\epsilon_{\mu\nu\lambda\rho\s}G^{\lambda\rho}V_{\s}^a+\ft{2}{\sqrt{6}}
e^{\fft1{\sqrt6}\phi}fF_{\mu\nu}^a\nonumber\\
&&\qquad-\ft1{\sqrt{6}}e^{\fft1{\sqrt6}\phi}\epsilon_{\mu\nu\lambda\rho\s}
F^{\lambda\rho\,a}K_{\s}
-\ft4{\sqrt{6}}e^{\fft1{\sqrt6}\phi}\Phi_{[\mu|\lambda|}^{ab}
F^{\lambda}}{_{\nu]}^{b},\\
0&=&\ft14{\epsilon_{\mu\nu}}^{\lambda\rho\s}\Phi_{\rho\sigma}^{ab}
\p_{\lambda}\phi
+\f{2}{\sqrt{6}}e^{-\fft2{\sqrt6}\phi}{\Phi_{[\mu}}^{\lambda\,ab}
{G^{\lambda}}_{\nu]}-\ft{2}{\sqrt{6}}e^{\fft1{\sqrt6}\phi}
f^{[a}F_{\mu\nu}^{b]}\nn\\
&&\qquad-\ft{1}{\sqrt{6}}e^{\fft1{\sqrt6}\phi}
{\epsilon_{\mu\nu}}^{\lambda\rho\s}
F_{\lambda\rho}^{[a}V_{\s}^{b]}+\ft1{\sqrt{6}}e^{\fft1{\sqrt6}\phi}
\epsilon^{abcde}
\Phi_{[\mu|\lambda|}^{cd}{F^{\lambda}}_{\nu]}^e.
\eea

The true differential identities are obtained by taking a covariant
derivative of the bilinears (\ref{eq:so5bilin}), and then using the
gravitino variation (\ref{eq:deltapsi}) to reexpress $\nabla_\mu\epsilon^i$
in terms of algebraic expressions.  The result is
\bea
{\na}_{\mu}f&=&-\ft13e^{-\fft2{\sqrt6}\phi}G_{\mu\nu}K^{\nu}
+\ft23e^{\fft1{\sqrt6}\phi}F_{\mu\nu}^aV^{\nu\,a},\\
{\na}_{\mu}f^a&=&-\ft13e^{-\fft2{\sqrt6}\phi}G_{\mu\nu}V^{\nu\,a}
+\ft{1}{12}e^{\fft1{\sqrt6}\phi}{\epsilon_{\mu}}^{\nu\lambda\rho\s}
F_{\nu\lambda}^b
\Phi_{\rho\s}^{ab}+\ft{2}{3}e^{\fft1{\sqrt6}\phi}F_{\mu\nu}^aK^{\nu},\\
\na_\mu K_\nu&=&\ft{1}{12}{\epsilon_{\mu\nu}}^{\rho\lambda\s}
e^{-\fft2{\sqrt6}\phi}
G_{\rho\lambda}K_{\s}+\ft{1}{3}e^{-\fft2{\sqrt6}\phi}fG_{\mu\nu}
-\ft16{\epsilon_{\mu\nu}}^{\rho\lambda\s}e^{\fft1{\sqrt6}\phi}F_{\rho\lambda}^a
V_{\s}^a\nonumber\\
&&\qquad-\ft23e^{\fft1{\sqrt6}\phi}F_{\mu\nu}^af^a,\\
\na_\mu V_\nu^a&=&\ft1{12}{\epsilon_{\mu\nu}}^{\rho\lambda\s}
e^{-\fft2{\sqrt6}\phi}
G_{\rho\lambda}V_{\s}^a+\ft13e^{-\fft2{\sqrt6}\phi}f^aG_{\mu\nu}
-\ft16{\epsilon_{\mu\nu}}^{\rho\lambda\s}e^{\fft1{\sqrt6}\phi}
F_{\rho\lambda}^aK_{\s}\nonumber\\
&&\qquad-\ft23e^{\fft1{\sqrt6}\phi}fF_{\mu\nu}^a
-\ft16e^{\fft1{\sqrt6}\phi}(g_{\mu\nu}\Phi_{\rho\s}^{ab}
-2g_{\nu\rho}{\Phi_{\mu\s}}^{ab}-4g_{\mu\rho}\Phi_{\nu\s}^{ab})F^{\rho\s\;b},\\
\label{eq:dPhi}
{\na}_{\mu}\Phi_{\nu\lambda}^{ab}&=&\ft1{12}\left(-g_{\mu[\nu}
{\epsilon{_\lambda]}^{\rho\s\a\b}}
+2\d _{[\nu}^{\rho}{\epsilon_{\lambda]\mu}}^{\s\a\b}-2\d_{\mu}^{\rho}
{\epsilon_{\nu\lambda}}^{\s\a\b}\right)\nonumber\\
&&\qquad\times\left(e^{-\fft2{\sqrt6}\phi}G_{\rho\s}
\Phi_{\a\b}^{ab}-e^{\fft1{\sqrt6}\phi}\epsilon^{abcde}
\Phi_{\a\b}^{cd}F_{\rho\s}^e\right)
-\ft{1}{3}{\epsilon_{\nu\lambda\mu}}^{\rho\s}e^{\fft1{\sqrt6}\phi}
f^{[a}F_{\rho\s}^{b]}\nonumber\\
&&\qquad+\ft23(-2g_{\mu[\nu}\d_{\lambda]}^{\rho}g^{\s\a}
-\d_{[\nu}^{\rho}\d_{\lambda]}^{\s}\d_{\mu}^{\a}
+4\d_{\mu}^{\rho}\d _{[\nu}^{\s}\d_{\lambda]}^{\a})e^{\fft1{\sqrt6}\phi}
V_{\a}^{[a}F_{\rho\s}^{b]}.
\eea

The above dilatino and gravitino equations simplify when combined.
Using a form notation, we first have the `zero form' expressions
\begin{eqnarray}
\label{eq:Kphi}
i_Kd\phi&=&0,\\
\label{eq:Vphi}
i_{V^a}d\phi&=&-\ft1{\sqrt{6}}e^{\fft1{\sqrt6}\phi}\Phi_{\mu\nu}^{ab}
F^{\mu\nu\,b},\\
0&=&\Phi_{\mu\nu}^{ab}G^{\mu\nu}+\ft1{2}e^{\fft3{\sqrt6}\phi}
\epsilon^{abcde}\Phi_{\mu\nu}^{cd}F^{\mu\nu\,e}.
\end{eqnarray}
The one-form differential identities are
\begin{eqnarray}
\label{KG}
d(e^{\fft2{\sqrt6}\phi}f)&=&i_KG,\\
\label{VF}
d(e^{-\fft1{\sqrt6}\phi}f)&=&-i_{V^a}F^a,\\
\label{KF}
d(e^{-\fft1{\sqrt6}\phi}f^a)&=&-i_KF^a,\\
\label{VG}
d(e^{\fft2{\sqrt6}\phi}f^a)&=&i_{V^a}G+e^{\fft3{\sqrt6}\phi}
*(\Phi^{ab}\wedge F^b),
\end{eqnarray}
while the two-form differential identities are
\begin{eqnarray}
\label{K*G}
d(e^{-\fft2{\sqrt6}\phi}K)&=&i_K(e^{-\fft4{\sqrt6}\phi}*G)
-2(e^{-\fft1{\sqrt6}\phi}f^a)F^a,\\
\label{V*F}
d(e^{\fft1{\sqrt6}\phi}K)&=&-i_{V^a}(e^{\fft2{\sqrt6}\phi}*F^a)
+e^{-\fft1{\sqrt6}\phi}fG-e^{\fft2{\sqrt6}\phi}f^aF^a,\\
\label{K*F}
d(e^{\fft1{\sqrt6}\phi}V^a)&=&-i_K(e^{\fft2{\sqrt6}\phi}*F^a)
+(e^{-\fft1{\sqrt6}\phi}f^a)G-e^{\fft2{\sqrt6}\phi}fF^a,\\
\label{V*G}
d(e^{-\fft2{\sqrt6}\phi}V^a)&=&i_{V^a}(e^{-\fft4{\sqrt6}\phi}*G)
-2e^{-\fft1{\sqrt6}\phi}fF^a+e^{-\fft1{\sqrt6}\phi}
\Phi_{\mu\lambda}^{ab}F^{\lambda\,b}_{\;\;\,\nu}dx^\mu\wedge dx^\nu.\qquad
\end{eqnarray}
In addition, the symmetrized rank two equations are
\begin{eqnarray}
\na _{(\mu}K_{\nu )}&=&0,
\label{eq:killing}\\
\na _{(\mu}V_{\nu )}^a&=&-\ft{1}{6}(g_{\mu\nu}\Phi
_{\rho\s}^{ab}-3g_{\mu\rho}\Phi _{\nu\s}^{ab}-3g_{\nu\rho}\Phi
_{\mu\s}^{ab})e^{\fft1{\sqrt6}\phi}F^{\rho\s\,b}.
\end{eqnarray}
In particular, we see that $K^\mu$ identically satisfies the Killing equation.
Finally, we may obtain expressions involving the two-form $\Phi^{ab}$
\begin{eqnarray}
d(e^{\fft3{\sqrt6}\phi}\Phi^{ab})&=&*\left[\left(e^{\fft1{\sqrt6}\phi}
\Phi^{ab}_{\mu\lambda}G^\lambda{}_\nu
+\ft1{2}e^{\fft4{\sqrt6}\phi}\epsilon^{abcde}\Phi^{cd}_{\mu\lambda}
F^{\lambda\,e}_{\;\;\nu}\right)dx^\mu\wedge dx^\nu-4e^{\fft4{\sqrt6}\phi}
f^{[a}F^{b]}\right],\nonumber\\
d(e^{-\fft1{\sqrt6}\phi}*\Phi^{ab})&=&-\ft1{2}\epsilon^{abcde}
\Phi^{cd}\wedge F^e.
\label{eq:d*Phi}
\end{eqnarray}
Equations (\ref{eq:Kphi}) through (\ref{eq:d*Phi}), along with the
covariant derivative on $\Phi_{\mu\nu}^{ab}$ given in (\ref{eq:dPhi})
form a complete set of differential identities.

Note that, by taking an exterior derivative of (\ref{KG}), (\ref{KF}),
(\ref{K*G}) and (\ref{K*F}), and by using the relation $\mathcal L=di_K+i_Kd$
for the Lie derivative, we may obtain
\bea
{\cal L}_K G&=&0,\\
{\cal L}_K F^a&=&0,\\
{\cal L}_K (e^{-\fft4{\sqrt6}\phi}*G)&=&i_K[d(e^{-\fft4{\sqrt6}\phi}*G)
-F^a\w F^a],\\
{\cal L}_K (e^{\fft2{\sqrt6}\phi}*F^a)&=&i_K[d(e^{\fft2{\sqrt6}\phi}*F^a)
-F^a\w G].
\eea
The last two lines vanish by the gauge field equations of motion.  These
expressions, along with (\ref{eq:Kphi}) and the Killing equation
(\ref{eq:killing}), ensure that the isometry generated by $K$ extends to
the entire solution.  Finally, using (\ref{K*F}) along with the Fierz
identity (\ref{eq:KVf}), we may also deduce that
\be
{\cal L}_K (e^{\fft1{\sqrt6}\phi}V^a)=0.
\ee
%


\end{document}